\documentclass[12pt,twoside]{article}
\usepackage{amsmath}
\usepackage{amssymb}
\usepackage{amscd}
\usepackage{epsfig}
\input xy
\xyoption{all}
\usepackage{epsfig}
\usepackage[latin1]{inputenc}
\def\be{\begin{equation}}
\def\ee{\end{equation}}
\def\ba{\begin{array}}
\def\bacc{\begin{array} {cc}}
\def\ea{\end{array}}
\def\bea{\begin{eqnarray}}
\def\eea{\end{eqnarray}}
\def\bd{\begin{displaymath}}
\def\ed{\end{displaymath}}

\def\D{\mathcal{D}}
\def\a{\alpha}
\def\b{\beta}

\def\Box{ {\,\lower 0.9pt\vbox{\hrule\hbox{\vrule
height0.2cm \hskip 0.2cm
\vrule height 0.2cm }\hrule}\,}}

\def\A{{\scriptscriptstyle A}}
\def\B{{\scriptscriptstyle B}}

\def\D{{\scriptscriptstyle D}}
\def\I{{\scriptscriptstyle I}}
\def\J{{\scriptscriptstyle J}}
\def\K{{\scriptscriptstyle K}}
\def\M{{\scriptscriptstyle M}}
\def\N{{\scriptscriptstyle N}}
\def\P{{\scriptscriptstyle P}}
\def\Q{{\scriptscriptstyle Q}}

\def\Lm{{\scriptscriptstyle \Lambda}}
\def\Gm{{\scriptscriptstyle \Gamma}}
\def\Om{{\scriptscriptstyle \Omega}}

\def\Tr{{\rm Tr}}
\def\exd{{\rm d}}

\begin{document}

\begin{center}

{\Large\bf The Fate of Unstable Gauge Flux Compactifications}

\vspace{1cm}

{\large C.P. Burgess$^{a,b}$\footnote{Email:
    cburgess@perimeterinstitute.ca},
  S.L. Parameswaran$^c$\footnote{Email:
    susha.louise.parameswaran@desy.de} and
I. Zavala$^d$\footnote{Email: zavala@th.physik.uni-bonn.de}}

\vspace{.6cm}

{\it {$^a$  Department of Physics and Astronomy, McMaster
    University,\\
1280 Main Street West, Hamilton ON, L8S 4M1, Canada.}}

\vspace{.4cm} {\it {$^b$  Perimeter Institute for
   Theoretical Physics,\\
31 Caroline Street North, Waterloo ON, N2L 2Y, Canada.}}

\vspace{.4cm}

{\it {$^c$  II. Institut für Theoretische Physik der Universität Hamburg,\\
DESY Theory Group, Notkestrasse 85, Bldg. 2a, D-22603 Hamburg,
Germany.}}

\vspace{.4cm}

{\it {$^d$ Bethe Center for Theoretical Physics and \\
Physikalisches Institut der Universität Bonn, \\
Nußallee 12, D-53115 Bonn, Germany.}}

\end{center}

\vspace{1cm}

\begin{abstract}

Fluxes are widely used to stabilise extra dimensions, but the
supporting monopole-like configurations are often unstable,
particularly if they arise as gauge flux within a non-abelian
gauge sector. We here seek the endpoint geometries to which this
instability leads, focussing on the simplest concrete examples:
sphere-monopole compactifications in six dimensions. Without
gravity most monopoles in non-abelian gauge groups are unstable,
decaying into the unique stable monopole in the same topological
class. We show that the same is true in Einstein-YM systems, with
the new twist that the decay leads to a shrinkage in the size of
the extra dimensions and curves the non-compact directions: in $D$
dimensions a $\hbox{Mink}_{D-2} \times S_2$ geometry supported by
an unstable monopole relaxes to $\hbox{AdS}_{D-2} \times S_2$,
with the endpoint sphere smaller than the initial one. For
supergravity the situation is more complicated because the dilaton
obstructs such a simple evolution. The endpoint instead acquires a
dilaton gradient, thereby breaking some of the spacetime
symmetries. For 6D supergravity we argue that it is the 4D
symmetries that break, and examine several candidates for the
endpoint geometry. By using the trick of dimensional oxidation it
is possible to recast the supergravity system as a
higher-dimensional Einstein-YM monopole, allowing understanding of
this system to guide us to the corresponding endpoint. The result
is a Kasner-like geometry conformal to $\hbox{Mink}_4 \times S_2$,
with nontrivial conformal factor and dilaton breaking the maximal
4D symmetry and generating a singularity. Yet the resulting
configuration has a lower potential energy than did the initial
one, and is perturbatively stable, making it a sensible candidate
endpoint for the evolution.

\end{abstract}

\newpage

\tableofcontents


\section{Introduction}

The ubiquity of moduli in extra-dimensional compactifications has
been a persistent thorn in the side of model-builders attempting
to bring higher-dimensional theories into contact with Nature as
we see it around us. For this reason flux-supported
compactifications, for which various $n$-form field strengths
thread cycles and are topologically blocked from relaxing to zero,
represent a significant step forward by providing an attractive
mechanism that dynamically stabilises many of these moduli.

Better yet, the required $n$-form fields are as common as dirt in
supersymmetric theories, arising as components of the gravity
supermultiplet in higher dimensions; as Maxwell fields required by
anomaly cancellation; or as fields sourced by D-branes or other
such objects. Perhaps the simplest such construction, due to Salam
and Sezgin \cite{SS}, is more than 20 years old, and threads a
Maxwell flux through the extra dimensions in 6D supergravity to
stabilise its compactification to $\hbox{Mink}_4 \times S_2$.

What is less well known is that a great many of such monopole
configurations are unstable, particularly when the flux involved
arises as a Dirac monopole embedded into a non-Abelian gauge
sector. For instance, explicit calculations \cite{prs2} show that
sphere-monopole compactifications in anomaly-free supergravity --
and their warped braneworld generalizations -- are generically
unstable, even though the monopole in question carries nontrivial
topological charge. The instability is possible because there are
typically more monopole solutions than there are distinct
topological sectors, allowing most to decay to the (often unique)
stable representative in any topological class --- a phenomenon
that is well understood within pure Yang Mills (YM) theories
\cite{brandtneri1,brandtneri2,coleman}.

For monopole-supported systems, the coupling to gravity does not
remove the instability \cite{seifinstabilities, schellekens}, and
requires the geometry also to relax as the monopole decays. We
examine this relaxation here, and argue that it is fairly
straightforward for the Einstein-YM system (EYM). As in pure YM
theory, the unstable monopole evolves towards the unique stable
monopole in the same topological class \cite{coleman}, and as it
does so the geometry adjusts simply by shrinking the size of the
supported extra-dimensional sphere, and by curving the large 4
dimensions. In $d+2$ dimensions, starting from $\hbox{Mink}_d
\times S_2$ the system evolves towards AdS${}_d \times S_2'$, with
the radius of $S'_2$ being smaller than that of $S_2$.

The situation is more complicated in the supergravity case,
because the dilaton obstructs this same simple evolution towards
another maximally symmetric solution built with the stable
monopole because for it $\Box \sigma \ne 0$. As a result
$\partial_\M \sigma \ne 0$, instead leading to a breakdown of some
of the spacetime symmetries. The corresponding final state for
higher dimensional systems, with gravity and a dilaton
back-reacting to the monopole dynamics, is unknown.

In this paper we examine several candidate stable endpoint
configurations for the simplest case of compactifications of 6D
supergravity down to 4D. We argue that the insensitivity of the
low-energy effective 4D scalar potential to scalar gradients in
the compactified two dimensions make it likely that it is the 4D
spacetime symmetries that break in this case, rather than those of
the compactified two dimensions.

To find the endpoint solution we employ a trick: a cycle of
dimensional oxidation and reduction that maps the solutions of the
supergravity of interest onto those of a dilaton-free pure EYM
system in still-higher dimensions. We use this to map the unstable
initial monopole-supported supergravity configuration onto an
unstable monopole-supported state in the still-higher dimensional
theory. Assuming this higher-dimensional EYM system relaxes in the
simple maximally-symmetric way tells us its endpoint, and this can
then be mapped to determine the endpoint EYM-dilaton configuration
that is supported by the final stable state into which the
monopole decays.

Proceeding in this way we are led to a stable, nonsupersymmetric
endpoint geometry that (in the absence of brane sources) is
conformal to $\hbox{Mink}_4 \times S_2$. Its nontrivial conformal
factor and dilaton break the maximal 4D symmetry, giving rise to a
singular geometry for which the dilaton and curvature blow up at a
point in the 4D spacetime. However, the configuration nonetheless
has a lower potential energy than did the initial one, and is
stable, and is a reasonable candidate for the endpoint of the
instability.

Although the dilaton changes the dynamics drastically, the
presence of branes (specifically, conical singularities in the
extra-dimensional geometry) and warping do not make much
difference, as we show by also finding a warped generalization of
the endpoint solution in this case, for generic brane tensions.
The solutions we find in this way turn out to be static analogues
of the time-dependent scaling solutions to 6D supergravity found
in \cite{scaling}, with the fields varying along a 4D spatial
coordinate rather than along time.

We also examine a class of supersymmetric solutions to 6D
supergravity as candidate endpoints (that also break the 4D
spacetime symmetry) \cite{glps}. Although we cannot prove these
not to be the ultimate endpoint, we provide arguments as to why
this seems less likely than those we construct using the
oxidation/reduction trick.

The rest of our exposition is organised as follows. The next
section, \S1, summarises the field equations of chiral gauged 6D
supergravity \cite{Nishino:1984gk}, together with their most
general monopole-supported solutions that have at most conical
singularities \cite{GGP,ocho}. This section concludes by briefly
summarising the linearised stability analysis of ref.~\cite{prs2},
and reviewing the topological classification of non-abelian Dirac
monopoles in YM theories. \S3 then describes how gravity
backreacts to monopole decay in dilaton-free EYM theory, by
shrinking the extra dimensions and curving the 4 large dimensions.
Finally \S4 generalises these considerations to the EYM-dilaton
system that arises in the 6D supergravity of interest. This
section describes the oxidation/reduction procedure, and applies
it to two examples. The first example considers unwarped systems
such as arise in the absence of branes, or with two branes having
equal tension. The second does the case of the general warped
geometries of \S2, having only conical singularities.  We end with
some brief conclusions.

\section{Theory and Background} \label{S:theory}

We start with the bosonic action\footnote{For fermionic terms see
\cite{Nishino:1984gk}.} for chiral 6D gauged supergravity coupled
to gauge- and hyper-multiplet matter, with gauge group
${\hat{\mathcal
G}} = {\mathcal{G}} \times U(1)_R$ \cite{Nishino:1984gk}
\bea
 S_B &=&\int d^6 x \sqrt{-g} \left[\frac{1}{\kappa^2} \, R
 -\frac{1}{4} \, \partial_\M\sigma \, \partial^\M \sigma
 -\frac{1}{4} \, e^{\kappa\sigma/2}\, \Tr \left( F_{\M\N}
 F^{\M\N}  \right)
 \right.\\
 &&\qquad \qquad \qquad
 -\frac{1}{12} \, e^{\kappa\sigma}
 H_{\M\N\P} H^{\M\N\P} \left.- G_{\a
 \b}(\Phi) \, D_\M \Phi^{\a} \, D^\M\Phi^{\beta}
 -\frac{8}{\kappa^4} \, e^{-\kappa
 \sigma/2}v(\Phi)\right] \,, \nonumber
 \label{SB}
\eea
where $\{ g_{\M\N}, H_3 = \exd B_2 + A_1 \wedge F_2, \sigma \}$ are the bosonic
fields in the gravity multiplet; $F_{\M\N}$ are the
gauge-multiplet field strengths for ${\mathcal{G}} \times
U(1)_R$; and $\Phi^\a$ denote the hyper-multiplet scalars. The
dependence of the scalar potential on $\Phi^{\a}$ is such that its
minimum is at $\Phi^{\a}=0$ where $v(0) = g_1^2$, and so we fix
henceforth $\Phi^{\a} =0$. Here $g_1$ is the $U(1)_R$ gauge
coupling, and we use $g$ for the ${\mathcal G}$
coupling constants.\footnote{In general, if ${\mathcal G}$
consists of several simple factors, ${g}$ represents a
collection of independent gauge couplings.}

Using $\Phi^\a = 0$ the remaining equations of motion (EOMs)
become
\bea
 &&\frac{1}{\kappa^2} \, R_{\M\N} = \frac{1}{4} \,
 \partial_{\M}\sigma \, \partial_\N \sigma +
 \frac{1}{2}\, e^{\kappa\sigma/2} \, \Tr \left(F_{\M\P}
 F^{\,\,\,\,\,\P}_{\N}\right)
 + \frac{1}{4} \,
 e^{\kappa\sigma} \, H_{\M\P\Q} H_\N^{\,\,\, \P\Q}
 -\frac{1}{4\kappa} \, g_{\M\N} \Box\sigma, \nonumber \\
 &&\frac{1}{\kappa} \, \Box\sigma=\frac{1}{4} \,
 e^{\kappa\sigma/2} \, \Tr \left(F_{\M\N} F^{\M\M}\right)
 + \frac{1}{6} \, e^{\kappa\sigma} H_{\M\N\P}H^{\M\N\P}
 -\frac{8g_1^2}{\kappa^4} \, e^{-\kappa\sigma/2},\nonumber \\
 && D_\M\left(e^{\kappa\sigma/2} F^{\M\N}\right)=\frac{\kappa}{2}
 \, e^{\kappa\sigma} \, H^{\N\P\Q} {F}_{\P\Q},\nonumber \\
 && D_\M\left(e^{\kappa\sigma} \, H^{\M\N\P}\right)=0,
\label{EOM}\eea
where
\bea
 && A_{\M} = A_{\M}^{\I}\, T_{\I} \,, \qquad   F_{\M\N} =
 F_{\M\N}^{\I}\,T_{\I}  \nonumber \\
 && F_{\M\N}^{\I} = \partial_{\M} A_\N^{\I} - \partial_\N
 A_\M^{\I} + g \, {c^\I}_{\J\K} A_\M^{\J}
 A_\N^{\K}            \nonumber  \\
 && D_\M = \nabla_\M \,\, - i  g  A_{\M}^{\I}\,T_{\I} \eea
with $\nabla_\M$ the Lorentz covariant derivative, and $T_{\I}$ are
the gauge group generators with structure constants ${c^{\I}}_{\J\K}$.

\subsection{Background solutions}

The solutions to these equations whose stability is of interest
are monopole-supported extra dimensions, in which the extra
dimensions are supported against gravitational collapse by having
a gauge flux thread the extra dimensions. Our interest in
particular lies in the case where this background flux lies within
the non-Abelian part of the gauge group. A broad class of these
have the form \cite{ocho},
\bea
 && \exd s^2 = g_{\M\N} \, \exd x^\M \exd x^\N = \rho \, \eta_{\mu\nu}
 \, \exd x^{\mu} \exd x^{\nu} +
 \frac{\exd\rho^2}{h(\rho)} + \, h(\rho) \, \exd\phi^2
 \nonumber \\
 && A_{\pm} = \frac{q^a \, Q_a}{2}
 \left(\frac{1}{\rho^2} - \frac{1}{\rho_{\pm}^2}\right)
 \exd\phi  \nonumber \\
 && \kappa \, \sigma = 2 \ln \rho \,,
 \qquad \qquad H_{\M\N\P} = 0
 \label{E:8author}
\eea
with
\bea \label{E:8author1}
 h(\rho) &=& \frac{2 M}{\rho} - \frac{2\, g_1^2 \rho}{\kappa^2} -
 \frac{\kappa^2 \, \gamma_{ab} q^a q^b}{8\,\rho^3}
 \nonumber\\
 &=& - \frac{2\, g_1^2}{\kappa^2 \rho^3}(\rho^2 - \rho^2_+)
 (\rho^2 - \rho^2_-) \,,
\eea
where $Q_a$ are the generators of the Cartan subalgebra of the Lie
algebra associated with the group ${\mathcal{G}}$,
normalized so that $\Tr(Q_a Q_b) = \gamma_{ab} = \gamma^2
\delta_{ab}$, for constant $\gamma$. The $q^a$ identify the
magnitude and direction of the background flux in the Lie algebra
of ${\mathcal{G}}$. Finally, $\rho_- < \rho < \rho_+$, where
\be
 \rho_\pm = \frac{\kappa^2}{2 g_1^2} \left[ M \pm
 \sqrt{M^2 - \frac14 \, \gamma_{ab} q^a q^b} \right] \,,
\ee
denote the two positive values where $h(\rho_\pm) = 0$, at which
point the geometry has a conical singularity, with deficit angle
\be
 \delta_{\pm} = 1- \frac12 \left|
 h'(\rho_{\pm}) \right|
 = 1 - \frac{2\, g_1^2}{\kappa^2 \rho_\pm} \, \left(
 \rho_+^2 - \rho_-^2 \right) \,.
\ee
As shown in ref.~\cite{bqtz}, it is the property that these
singularities are conical that defines these solutions,
eqs.~(\ref{E:8author}) and (\ref{E:8author1}), as special cases of
the more general solutions of ref.~\cite{GGP}.

We regard the conical singularities as indicating the presence of
source codimension-two branes having tensions $T_{\pm}$,
\be
 S_{brane} = -T_{\pm} \int d^4y \sqrt{-\gamma_{\pm}} \,,
\ee
with $y^i$ being coordinates on the brane world-volume, and
$\gamma_{ij}$ the induced metric there. The tension is related to
the geometry's conical defect angle through $T_{\pm} =
2\delta_{\pm}/\kappa^2$, and this connection allows us to trade
the integration parameters $M$ and $q^2 = \gamma_{ab} q^a q^b$ for
the two source brane tensions.  It turns out that only one
combination of these parameters is fixed, and that the tensions of
the branes are related to each other by a constraint \cite{ocho}
(see later sections for a recap of some of these features).

These solutions break supersymmetry, apart from the special
rugby-ball case where the dilaton is a constant: $\partial_m
\sigma = 0$. Whether supersymmetry breaks even in this case
depends on the boundary conditions at the branes \cite{hm}, which
is governed by more model-dependent details of the branes
themselves.

The amplitude, $q^a$, of the gauge field is also constrained by
topology to be quantized, as follows. In order for the gauge field
potential to be well-defined at $\rho_{\pm}$, we need to cover the
internal manifold with two coordinate patches. Requiring that the
two patches be related by a single valued gauge transformation on
their overlap leads to the following Dirac quantization condition
\be \label{diracquantization1}
 - g \, e_{a\I} \, \frac{q^a}{2}
 \left(\frac{1}{\rho_+^2} -
  \frac{1}{\rho_-^2}\right) = N_{\I} \,,
\ee
where $N_{\I}$ are integer {\em monopole numbers}, one for each
gauge generator $T_\I$. The quantities $e_{a\I}$ denote the $Q_a$
charge of generator $T_\I$, defined in the adjoint representation
by choosing a basis of generators that satisfies $\left[ Q_a, T_\I
\right] = e_{a\I} T_\I$ (no sum). This clearly vanishes for all
generators of the Cartan subalgebra, $e_{ab} = - e_{ba} = 0$. For
those $T_i$ not in the Cartan subalgebra\footnote{In the
Cartan-Weyl basis of generators we label the Lie algebra of
${\mathcal{G}}$ by: $\{T_\I \} = \{Q_a, T_i , T_{-i}\}$.}
hermitian conjugation reverses the sign of this charge, so we
choose notation so that $T_i^\dagger = T_{-i}$.

\medskip\noindent{\em Rugby Balls and Spheres:}

\medskip\noindent
Ref.~\cite{bqtz} shows that these solutions go over to the
unwarped rugby-ball solutions \cite{abpq}, when $T_- \rightarrow
T_+$ (and to the spherical Salam-Sezgin solutions \cite{SS} when
$T_+ = T_- = 0$). In these limits, a change of coordinates puts
the background into the familiar form of a spherical geometry
supported by a Dirac monopole:
\bea
 && \exd s^2 = \eta_{\mu\nu} \exd x^{\mu} \exd x^{\nu}
 + a^2 \left(\exd\theta^2 +
  \sin^2\theta \exd\phi^2 \right) \nonumber \\
 && A_\pm = -\frac{q^a Q_a}{2}
 \left(\cos\theta \mp 1 \right) \exd\phi
 \nonumber \\
 &&  \kappa\sigma =  H_{\M\N\P} = 0
 \label{E:sphere}
\eea
In this case the equations of motion fix the radius of the
sphere,\footnote{More generally there is a flat direction along
which the values of $\sigma$ and $a$ are correlated.} $a =\kappa /
(2{\sqrt 2} g_1)$, and fix the monopole strength
\be \label{rugbymonopole}
 q^2 = \gamma^2 \delta_{ab} \, q^a q^b  = \frac{1}{g_1^2} \,.
\ee
(Any other value for the monopole strength would induce warping in
the non-compact directions, which requires $T_+ \ne T_-$). On the
other hand, the Dirac quantization condition in this case reduces
to:
\be \label{diracquantization2}
 - {g} \, q^a \, e_{a\I} = N_{\I} \,,
\ee
and so consistency between this and eq.~(\ref{rugbymonopole}) in
general requires relations between the otherwise independent
couplings $g_1$ and $g$. In the simplest case where the
monopole is aligned in the $U(1)_R$ direction \cite{SS} we have
$g = g_1$ and consistency between equations of motion and
Dirac quantisation imply the monopole number must be $N = \pm 1$.

\subsubsection*{A concrete example}

It is useful in what follows to have in mind a concrete example
that is simple enough to solve explicitly yet complicated enough
to display the instabilities of later interest. For this purpose
we focus on the subsector of the theory for which the gauge fields
lie within a subgroup ${\hat{\mathcal G}} = SU(3) \times U(1)_R$ of the
full group, with all hyper-scalars either neutral under the
non-Abelian subgroup or transforming in the
adjoint,\footnote{Although the hyper-scalars vanish in the
background, the charge of their fluctuations plays a role in the
Dirac quantization conditions.}
 and all other fields required
for anomaly cancellation, including the Kalb-Ramond fields
$H_{\M\N\P}$, set to zero.

The Cartan subalgebra of $SU(3)$ is two-dimensional, $Q_a$ with
$a={\it 1, 2}$, and with the normalisation condition $\gamma_{ab}
= \Tr\left( Q_a Q_b \right) = \frac16 \, \delta_{ab}$ (so
$\gamma^2 = \frac16$), these may be written
\be
 Q_{\it 1} = \frac{1}{2\sqrt{3}} \left(
 \begin{array}{ccc}
  1 &  &  \\
   & -1 &  \\
   &  & 0 \\
 \end{array} \right) \qquad
 Q_{\it 2} = \frac{1}{6} \left(
 \begin{array}{ccc}
  1 &  &  \\
   & 1 &  \\
   &  & -2 \\
 \end{array} \right) \,.
\ee
The remaining six generators can be divided into three pairs,
$T_i$ and $T_{-i}$ with $i = 1,2,3$, having opposite charges. The
independent nonzero charge eigenvalues, $e_{ai}$, then become
\begin{figure}[ht]
\begin{center}
\begin{tabular}{c||ccc}
  & $T_1$ & $T_2$ & $T_3$ \\
  \hline\hline
  $Q_{\it 1}$ & $\frac{1}{\sqrt3}$ & $\frac{1}{2\sqrt3} $
  & $-\frac{1}{2\sqrt3} $ \\
  $Q_{\it 2}$ & 0 & $\frac12$ & $\frac12$
\end{tabular}\\
\vspace{0.3cm}{{\bf Table 1:} Table of charges for adjoint fields
in $SU(3)$.} \label{Tab:charges}
\end{center}
\end{figure}

The monopole breaks the $SU(3)$ gauge group down to either $U(1)_1
\times U(1)_2$ or $SU(2) \times U(1)_2$, depending on whether or
not all of the eigenvalues of $q^a Q_a$ are distinct or if two of
them are equal.

\medskip\noindent{\em The case $SU(3) \to SU(2) \times U(1)_2$:}

\smallskip\noindent
If two eigenvalues of $q^a Q_a$ are equal then an $SU(3)$ rotation
can be performed to ensure that $q^a Q_a$ points purely in the
$q^{\it 2}$ direction.\footnote{The same can sometimes also be
done if its eigenvalues all differ, but the required $SU(3)$
transformation is then singular, a distinction that turns out not
to be important for identifying which monopoles are topologically
stable \cite{brandtneri2}.} The spectrum of $SU(3)$ gauge bosons
then decomposes into the four massless gauge fields of the
unbroken gauge group together with an $SU(2)$ doublet of massive
charged states, having charge $\frac{1}{2}$ with respect to
$U(1)_2$ (and their conjugates). The Dirac quantisation condition
then requires that $q^{\it 2} = 2N/g$ where $N = N_{i=2} =
N_{i=3}$ is an arbitrary integer, while $N_a = N_{i=1} = 0$.

\medskip\noindent{\em The case $SU(3) \to U(1)_1 \times U(1)_2$:}

\smallskip\noindent
Alternatively, if all eigenvalues of $q^a Q_a$ are distinct then
both $q^{\it 1}$ and $q^{\it 2}$ are nonzero. The $SU(3)$ gauge
fields then decompose into two massless gauge fields, together
with three complex massive vectors with $U(1)_1 \times U(1)_2$
charges as given by Table 1: $(\frac{1}{ \sqrt{3}}, 0)$,
$(\frac{1}{ 2\sqrt{3}}, \frac12)$, $(-\frac{1}{ 2\sqrt{3}},
\frac12)$. The Dirac quantisation condition then requires that
$q^{\it 1} = {2\sqrt{3}} \, s^{\it 1}/{g}$ and $q^{\it 2} =
{2}\, s^{\it 2}/{g}$, where $s^{\it 1} = \frac12
(N_{i=3}-N_{i=2}) = \frac12 N_{i=1}$ and $s^{\it 2} =
\frac12(N_{i=3} + N_{i=2})$ are half-integer valued.

\medskip
Different quantum numbers $(s^{\it 1}, s^{\it 2})$ do not always label
distinct monopoles.
For
instance if $(s^{\it 1}, s^{\it 2}) = \left( \frac12, \frac12
\right)$, then
\be
 g \, q^a Q_a = \sqrt3 \, Q_{\it 1} + Q_{\it 2}
 = \frac13
 \left( \begin{array}{ccc}
  2 & & \\
  & -1 & \\
  & & -1 \\
 \end{array}\right) \,,
\ee
and so equals $-2 Q_{\it 2}$ up to a permutation of the axes. This
shows that the $(s^{\it 1}, s^{\it 2}) = \left( \frac12, \frac12
\right)$ monopole is physically equivalent to the $(s^{\it 1},
s^{\it 2}) = \left( 0, -1 \right)$ (or $N=-1$) $SU(2) \times
U(1)_2$-preserving monopole.

\subsection{Linearised instability} \label{S:instability}

Linearised stability analysis shows that spacetimes stabilised by
monopoles embedded into non-abelian groups (as above) are
unstable, as we now summarise following ref.~\cite{prs2}. Consider
therefore linearising about the background geometry
\be
 \bar g_{\M\N} \exd x^\M \exd x^\N :=
 e^{\bar A} \eta_{\mu\nu} \exd x^{\mu}
 \exd x^{\nu} + e^{\bar A} \exd u^2
 + e^{\bar B} \exd\phi^2  \,,
\ee
where the extra-dimensional coordinates are $\{ x^m \} = \{
u(\rho), \phi \}$. Denote the Ricci tensor for this geometry by
$\overline R_{\M\N}$, and the background gauge field by ${\cal
A}_\M$, with field strength ${\cal F}_{\M\N}$.

The unstable tachyonic directions turn out to be among the
Kaluza-Klein (KK) modes of the non-abelian gauge field that live
in the extra dimensions and lie along directions of the gauge
algebra that are charged under the generator along which the
background monopole points:
\bea
 \delta A_{u}^{\I} T_{\I} &:=& V_{u}^{\I} T_{\I}
 = V_{u} \nonumber \\
 \delta A_{\phi}^{\I} T_{\I} &:=& V_{\phi}^{\I} T_{\I}
 = V_{\phi}
 \,.
\eea
Raising and lowering all indices with the rescaled background metric,
$\hat g_{MN} = e^{\kappa\sigma/2} \, \overline g_{\M\N}$, ref.~\cite{prs2} shows that the part of the
action that is bilinear in these unstable gauge-field fluctuations
is (in light-cone gauge):
\bea
 S_{2}(V,V) &=& -\frac12 \int d^6X \sqrt{-\hat g} \; \Tr
 \left[\partial_{\mu} V_m \partial^{\mu} V^m + D_m V_n D^m V^n - 2
  (\partial_u \hat A)^2 V_u^2  \right. \nonumber \\
 && \phantom{000000000}  \left.
 - 2 (\partial_u \hat A) V_u D_mV^m
 + \hat R_{mn} V^m V^n + 2 {g} \,
 {\cal F}_{mn} V^m \times V^n \right] \, , \label{b2action}
\eea
where $\hat A = \bar A + \kappa\sigma/2$, and the covariant
derivative of $V_\M$ is defined by
\be
 D_{\M} V_\N = {\nabla}_{\M} V_\N - i g\, [{\cal
 A}_{\M}, V_\N] \,,
\ee
and ${\nabla}_{\M}$ is the Lorentz covariant derivative.

Solving the linearised equations of motion and boundary conditions obtained from this  action, and requiring the resulting modes to have finite kinetic energy, leads to a discrete spectrum of fluctuations. Taking advantage of the axial-symmetry, make the Fourier
decomposition:
\be
 V_n(X) = \sum_{\bf m} V_{n{\bf m}}(x,u) e^{i{\bf m}\phi}
 \label{fourier}
\ee
with ${\bf m}$ an arbitrary integer, $-\infty < {\bf m} < \infty$.
To diagonalise the mode functions make the field redefinitions
\be
 V_{\pm \, {\bf m}}(x,u) = \frac{1}{\sqrt{2}}\left(e^{(\hat A +
  \hat B)/4} \, V_{u \, {\bf m}}(x,u) \pm i
 e^{(3\hat A - \hat B)/4} \, V_{\phi \, {\bf m}}(x,u)\right)
\ee
and perform a Kaluza-Klein decomposition
\be
 V_{\pm}(x,u) = V_{\pm}(x) \psi_{\pm}(u) \,.
\ee

The solutions for $\psi_{\pm}(\rho)$ can then be found explicitly
in terms of hypergeometric functions. For ${\bf n}=0,1,2,\dots$
the corresponding KK mass spectrum for $V_+^{\I}$ is
\begin{itemize}
\item For $ {\bf m} \leq -\frac{1}{\eta_+}$ and ${\bf m} \leq N_{\I} +
  \frac{1}{\eta_-}$
\be
 M^2=\frac{1}{a^2}\left\{{\bf n}({\bf n}+1)-\left({\bf n}
 +\frac{1}{2}\right)\left[{\bf m} \eta_+ +({\bf m}-N_{\I})
 \eta_- \right] + {\bf m} ({\bf m}-N_{\I}) \eta_+ \eta_-
 \right\}.\label{M1}
\ee
\item For $-\frac{1}{\eta_+} < {\bf m} \leq N_{\I} +
  \frac{1}{\eta_-}  $
\be
 M^2=\frac{1}{a^2}\left\{\left({\bf n}+\frac32\right)^2
 - \frac14 +\left({\bf n}+\frac{3}{2}\right)
 \left[ {\bf m} \eta_+ -({\bf m}-N_{\I}) \eta_-
 \right]\right\}. \phantom{0000000000} \label{M2}
\ee
\item For $N_{\I} + \frac{1}{\eta_-}< {\bf m} \leq -\frac{1}{\eta_+}$
\be
 M^2=\frac{1}{a^2}\left\{{\bf n}({\bf n}-1)
 -\left({\bf n}-\frac{1}{2}\right)\left[
 {\bf m} \eta_+ - ({\bf m}-N_{\I}) \eta_-
 \right]\right\} \,.
 \phantom{000000000000000}
 \label{M3}
\ee
\item For ${\bf m}>-\frac{1}{\eta_+}$ and ${\bf m} > N_{\I} + \frac{1}{\eta_-}$
\be
 M^2=\frac{1}{a^2}\left\{{\bf n}({\bf n}+1)
 +\left({\bf n}+\frac{1}{2}\right)\left[
 {\bf m} \eta_+ +({\bf m}-N_{\I}) \eta_- \right]
 + {\bf m}({\bf m}-N_{\I}) \eta_+ \eta_-
 \right\}.\label{M4}
\ee
\end{itemize}
In these expressions the parameters, $\eta_\pm$, are related to
the two conical defect angles, $\delta_\pm$, by $\eta_\pm = \left(
1 - \delta_\pm/2\pi \right)^{-1}$. The spectrum for $V_-^{\I}$ is
obtained from the above by transforming ${\bf m} \rightarrow -{\bf
m}$ and $N_{\I} \rightarrow - N_{\I}$. The integer $N_\I$ is the
quantity appearing in the Dirac quantisation condition,
eq.~(\ref{diracquantization1}).\footnote{In the sphere limit, the
spectrum can be put into the familiar form $M^2 =
\frac{1}{a^2}\left[l(l+1) - \left(\frac{N_{\I}}{2}\right)^2
\right]$ with multiplicity $2l+1$, where, for $V^{\I}_{\pm}$, $l =
k + |1 \pm N_{\I}/2|$, and $k=0,1,2,\dots$.}

Using these expressions it is possible to show that the necessary
and sufficient condition for the absence of tachyonic modes,   
assuming brane tensions not less than zero,
is
\be
 \left|N_{\I}\right|\leq 1 \quad \mbox{for every}
 \,\,{I}.\label{spherecondition}
\ee
For fields with $N_{\I} \geq 2$ the tachyonic modes are those with
${\bf m}=1,2, \dots, N_{\I}-1$, while for fields $N_{\I}\leq -2$
the tachyonic modes are those with ${\bf m}=-1,-2, \dots,
N_{\I}+1$. Notice that because ${\bf m} \ne 0$ the instability
towards growth of these modes spontaneously breaks the axially
symmetry of the background.

In fact, it has long been known that non-abelian monopoles in pure
gauge theory in 4 dimensions are also generically unstable, with
only one dynamically stable monopole existing within each
topological class \cite{brandtneri1, coleman}. Similar
instabilities were also found soon after in higher dimensions,
compactified on spheres, both for Yang-Mills and for
Einstein-Yang-Mills theories \cite{seifinstabilities,schellekens}.
We use these related instabilities in subsequent sections to try
to identify the new state towards which the system evolves once
the instability develops.

\subsubsection*{The $SU(3) \times U(1)_R$ example}

It is instructive to apply this to the specific example considered
above, where the monopole is embedded into $SU(3) \times U(1)_R$.

\medskip\noindent{\em The case $SU(3) \to SU(2) \times U(1)_1$:}

\smallskip\noindent
In this case we had $(q^{\it 1},q^{\it 2}) = (2/g)(0,N)$,
with $N_{a} = N_{i=1} = 0$ and $N_{i=2}=N_{i=3} = N$. In this case
we find no tachyonic modes when $N = 0, \pm 1$, but instability
when $|N| \ge 2$ (for which there are two complex tachyonic modes,
$V^{i=2}$ and $V^{i=3}$).

\medskip\noindent{\em The case $SU(3) \to U(1)_1 \times U(1)_2$:}

\smallskip\noindent
Consider the monopole with $(s^{\it 1}, s^{\it 2})=(1,0)$, for
example, where solving for the $N_\I$'s leads to the nonzero
values $N_{i} = (2,1,-1)$. Since $N_{i=1} = 2$ this monopole has
one complex unstable tachyonic direction. Similarly, the monopole
with $(s^{\it 1}, s^{\it 2})=(2,0)$ has tachyons amongst all three
of its charged fluctuations, since $N_{i} = (4,2,-2)$. However,
the embedding $(s^{\it 1},s^{\it 2}) = \left( \frac12,\frac12
\right)$ turns out to give monopole numbers $N_{i} = (1,1,0)$, and
so is stable, as it must be given that it is equivalent to the
$N=-1$ monopole of the $SU(2) \times U(1)_2$-preserving category.
Since this (and its conjugate) is the only stable monopole of this
category, we see that the only three stable cases preserve $SU(2)
\times U(1)_2$, with $N= 0,\pm 1$.

\subsection{Topology}

Since topological charge can cause stability for some
configurations, it is worth identifying how these charges are
classified for non-abelian monopoles. In the present instance the
topology of the internal manifold is that of a sphere, with Euler
number $\chi=2$. This is not changed by the presence of the
conical defects, since the contribution to $\chi$ from the
singularities compensates for the reduction that the angular
defect causes in the contribution to $\chi$ from the integral of
$R$ over the internal space.

\begin{table}[h]
\begin{center}
\begin{tabular}{c|c|c}
Gauge group ${\mathcal G}$ & $\pi_1({\mathcal G})$  & Center
\\ \hline
 $SU(N)$ &  1 & $Z_N$ \\
 $SO(2k+1)$ & $Z_2$ & $Z_2$ \\
 $SO(4k)$ & $Z_2$ & $Z_2 \times Z_2$  \\
 $SO(4k+2)$ & $Z$ ($k=0$), $Z_2$ ($k \geq 1$) & $Z_4$  \\
$Sp(N)$ & 1 & $Z_2$
\\
 $E_8$ & 1 & 1
 \\
 $E_7$ & 1 & $Z_2$\\
$E_6$ & 1 & $Z_3$\\
$F_4$ & 1 & 1 \\
$G_2$ & 1 & 1
\end{tabular}
\end{center}
\caption{$\pi_1$ and centres for the simple
Lie groups. \label{lietopologies}}
\end{table}

The non-trivial topology associated with the Dirac monopole
embedded into the gauge group ${\mathcal G}$ is similarly
classified by $\pi_1({\mathcal G})$. This can be seen explicitly
in the so called Wu-Yang construction \cite{wuyang}, used above, wherein the
extra dimensions are covered with two patches, each with a
non-singular gauge potential $A_{\pm}$. In this case the $A_\pm$
must differ on the overlap of the two patches at the equator by a
single-valued gauge transformation, and so defines a map from
$S^1$ to ${\mathcal G}$ that is classified by $\pi_1({\mathcal
G})$.\footnote{If Higgs fields spontaneously break ${\mathcal G}
\rightarrow {\mathcal H}$, then magnetic charge would instead be
classified by $\pi_1({\mathcal H})$. For smooth configurations
without Dirac strings this reduces to the subgroup of
$\pi_1({\mathcal H})$ consisting of those elements which are
contractible in ${\mathcal G}$, denoted $\pi_1({\mathcal
H})_{\mathcal G}$. This is equivalent to the familiar
classification of non-singular monopoles by $\pi_2({\mathcal
G}/{\mathcal H})$ (interpreted as classifying the map from the
$S_2$ at spatial infinity to the vacuum manifold ${\mathcal
G}/{\mathcal H}$) due to the isomorphism $\pi_1({\mathcal
H})_{\mathcal G} \sim \pi_2({\mathcal G}/{\mathcal H})$.
Similarly, in the absence of a Higgs contribution to topological
charge the classification $\pi_1({\mathcal H})$ reduces to the
Wu-Yang classification $\pi_1({\mathcal G})$, since
$\pi_1({\mathcal H})/\pi_1({\mathcal H})_{\mathcal G} \sim
\pi_1({\mathcal G})$.}

For non-abelian groups the integer corresponding to this
topological classification can be written explicitly in terms of the
representative gauge fields. Suppose for example, the gauge
algebra is $SU(N)$, and all charged fields transform in the
adjoint representation. Then the global group is actually
${\mathcal G} = SU(N)/{\mathbb Z}_N$ because the adjoint
representation uses the same matrix to represent two group
elements that differ only by an element of the group's center.
Define the magnetic flux, $\Phi$, using the following integral,
\be
 \Phi = \frac{1}{N} \Tr \, \exp \left[ i g\oint \exd s
 \left( A_{+\M} - A_{-\M}
   \right) \frac{\exd x^{\M}}{\exd s} \right] \,,
\ee
where the path is taken as the closed loop around the equator in
the overlap of the two patches on which the two gauge
configurations, $A_+$ and $A_-$, are respectively defined. This
expression, when evaluated using explicit gauge configurations,
produces a phase
\be
    \Phi = \exp \left( \frac{2\pi i L}{N} \right) \,,
\ee
where $0 \le L < N$ is the integer that labels the corresponding
element of $\pi_1(SU(N)/Z_N) = Z_N$.

Table \ref{lietopologies} gives $\pi_1$ and the centers of all the
simple Lie algebras. Amongst the known anomaly-free non-abelian
gauge groups in 6D chiral supergravity, those involving
non-Abelian gauge groups with non-trivial topology are the classic
$E_7 \times E_6 \times U(1)_R$ model \cite{e6e7u1}, for which all
hyper-multiplets are singlets under $E_6$; as well as two models
by Avramis and Kehaghias \cite{alex}, respectively involving $E_6$
(with hyper-multiplets only in the adjoint representation) or
$SO(N)$'s.

\subsubsection*{The example $\mathcal{G} = SU(3) \times U(1)_R$}

For the Dirac monopole embedded in $SU(3)$, with all fields
charged under the $SU(3)$ subgroup transforming in the adjoint
representation, as described above the global group is actually
${\hat {\mathcal G}} = \left[ SU(3)/{\mathbb Z}_3\right] \times U(1)_R$,
and the topological classification is given by:
\bea
 \pi_1({\hat{\mathcal G}}) &=& \pi_1({{\mathcal G}})
 + \pi_1(U(1)_R)
 \nonumber \\
 &=& \pi_1(SU(3)/{\mathbb Z}_3) +
 \pi_1(U(1)_R) \nonumber \\
 &=& {\mathbb Z}_3 +{\mathbb Z}
\eea
Notice that each simple factor of ${\hat{\mathcal G}}$ gives rise to its
own topological classification, provided that the monopole lies at
least partially in the factor of interest.

If the monopole is embedded purely within the $SU(3)$ then the
topology is simply classified by $\pi_1(SU(3)/{\mathbb Z}_3) =
{\mathbb Z}_3$. In the case where the monopole preserves an
unbroken $SU(2) \times U(1)_2$ we have seen that it can always be
written as
\footnote{More generally, a monopole in $SU(3)$ can always be written
  as $g A_\pm = -\frac12 \, {\mathbb M} \, (\cos\theta \mp 1) \, \exd\phi$ where
  ${\mathbb M} = 2L \, Q_{\it 2} + diag(r_1,r_2,r_3)$, $L$ is one of
  $0,1,2$ and
  $r_1,r_2,r_3$ are integers that sum to zero \cite{brandtneri2}.}
\bea
 g A_\pm &=& -\frac12 \, {\mathbb M} \, (\cos\theta \mp 1)
 \, \exd\phi \nonumber \\
 \hbox{where} \qquad
 {\mathbb M} &=& 2 N \, Q_{\it 2} = \frac{N}{3}
 \left(\begin{array}{ccc}
  1 & & \\
  & 1 & \\
  & & -2 \\
\end{array} \right) \,,
\eea
and, as before, $Q_{\it 2}$ is the second Cartan generator of
$SU(3)$ while $N$ is the integer of the Dirac quantisation
condition. In this case the associated flux evaluates to $\Phi =
e^{i2\pi N/3} \in {\mathbb Z}_3$, showing that it is $L = N$ (mod
3) that labels the distinct topological class \cite{brandtneri2}.
The finding that linearised stability requires $N = 0,\pm 1$ is
therefore consistent with the expectation that there is only one
stable monopole in each of the three independent topological
sectors.

\section{The Instability's Endpoint}

The previous sections argue that monopole-supported flux
compactifications in 6D supergravity are generically unstable,
provided the monopole is embedded within a non-Abelian factor of
the gauge group. We now ask what the new configuration is towards
which such an unstable non-Abelian monopole evolves.

As mentioned above, this problem is well understood in the case of
pure Yang-Mills (YM) theory, where the instability describes the
decay into the lightest monopole within the given topological
class \cite{coleman}. Our goal is to address what such a decay
implies when the monopole in question supports an
extra-dimensional compactification. We do so in this section
starting with simple spherical compactifications of
extra-dimensional Yang Mills and Einstein-Yang-Mills (EYM) systems
(including a cosmological constant, $\Lambda$). We defer the
qualitatively different case of EYM-Dilaton theories relevant to
higher-dimensional supergravity to the next section.

Concretely, consider the unstable $N = 2$ $SU(3)$ monopole
described in previous sections, for which $ g (q^{\it 1},
q^{\it 2}) = (0,4)$. Because $N \ne 0$ (mod 3), this state has a
non-trivial topology that prevents it from decaying into a
topologically trivial configuration. It must instead decay into
the stable monopole with $N=-1$, doing so by emitting magnetic
radiation (see e.g. \cite{brandtneri1,coleman}).


\subsection{Einstein Yang-Mills Theory}

When the decaying monopole supports a compactified extra
dimension, its decay should also cause the extra-dimensional
geometry to change. But since the decay also reduces the 4D
monopole energy density, its decay should also change the
curvature of the large dimensions we observe. We first show how
this takes place in detail, working within the Einstein-Yang-Mills
system. In this case, we must solve both the Einstein and Maxwell
equations of motion to check that the expected stable monopole is
a possible endpoint solution.

Consider then six-dimensional gravity coupled to a Yang-Mills
field and a positive 6D cosmological constant, $\Lambda$. We start
with a solution, $\hbox{Mink}_4 \times S_2$, for this system
comprising an unstable $SU(3)$ monopole supporting two spherical
extra dimensions, with $\Lambda$ adjusted to allow the observable
four dimensions to be flat. This initial monopole then decays into
the topologically connected stable monopole as above, whilst the
background geometry appropriately adjusts itself. We do not try to
follow the time-dependence of this process in its full transient
glory. Instead we directly seek the endpoint solution to which it
ultimately evolves, under the assumption that this endpoint also
remains maximally symmetric in the 4 visible and 2 internal
dimensions.

It turns out that we are led in this way to two possible endpoint
solutions. Either the internal sphere shrinks whilst the
non-compact directions curve into anti-de Sitter space (AdS), or
the sphere grows whilst the 4D spacetime curves into a de Sitter
(dS) universe. Based on the principle that the evolution lowers
the effective 4D scalar potential energy, we expect it is the AdS
solution towards which the system evolves.

\subsubsection*{Endpoint Solutions}

We start with the 6D EYM action,
\be
 S_{EYM} = \frac{1}{\kappa^2}\int d^6 x \sqrt{-g} \left[R
 -\frac{\kappa^2}{4} \,  \Tr \left( F_{\M\N}
 F^{\M\N}  \right)
 - \Lambda \right] \,,
 \label{SBEYM}
\ee
containing only gravity, $g_{\M\N}$, the Yang-Mills field,
$A_{\M}$, and the 6D positive cosmological constant, $\Lambda$.
The equations of motion for this system become
\bea
 &&  R_{\M\N}
 = \frac{\kappa^2}{2} \Tr \left( F_{\M\P} F_\N^{\,\,\,\P}
 \right) + g_{\M\N} \left( \frac{\Lambda}{4} - \frac{\kappa^2}{16}
 \Tr F^2 \right)  \\
 && \nabla_\M F^{\M\N} - i {g} A_\M
 F^{\M\N} = 0 \,,
\eea
for which we seek solutions having maximal symmetry in both 4
large dimensions and 2 small ones
\be
 \exd s^2 = g_{\mu\nu} \exd x^\mu \exd x^\nu
 + a^2 (\exd\theta^2 + \sin^2\theta \exd\phi^2)
\ee
with the maximally symmetric metric, $g_{\mu\nu}$, satisfying
$R_{\mu\nu} = 3 \lambda g_{\mu\nu}$, with 4D curvature constant
$\lambda$. In our conventions the cases $\lambda > 0$, $\lambda =
0$ or $\lambda < 0$ respectively correspond to dS, flat and AdS
geometries.

Consider now solutions for which the Maxwell field strength only
has nonzero internal components, $F_{mn}$, and depends only on the
coordinate $\theta$ (as in the monopole solution). The equation of
motion for the gauge field then implies
\be
 F_{\theta\phi} = \frac{q^a Q_a}{2} \, \sin{\theta}
\ee
where the constants $q^a$ again parametrise the monopole strength.
Recalling our convention $\Tr(Q_a Q_b) = \gamma_{ab} = \gamma^2
\delta_{ab}$, we find after use of the 4D components of Einstein's
equations,
\be
 R_{\mu\nu} = 3\lambda g_{\mu\nu} = g_{\mu\nu} \left(
 \frac{\Lambda}{4} - \frac{\kappa^2 q^2}{32 \, a^4} \right) \,,
 \label{Rmunu}
\ee
where $q^2 = \gamma_{ab} q^a q^b$. From the 2D Einstein equations
we instead find
\be
 R_{mn} = \frac{g_{mn}}{a^2} =  g_{mn}\left[ \frac{
    3\, \kappa^2 q^2}{32 \,a^4} + \frac{\Lambda}{4} \right] \,,
 \label{Rmn}
\ee
leading to the following conditions for $\lambda$ and $a$:
\bea
 && 3\lambda = \left( \frac{\Lambda}{4}
 - \frac{\kappa^2 q^2}{32\,a^4} \right) \label{condi1} \\
 && \frac{1}{a^2} =  \frac{
    3 \, \kappa^2 q^2}{32 \,a^4} + \frac{\Lambda}{4}
    = \frac{\kappa^2 q^2}{8\,a^4}
 + 3\lambda \,. \label{condi2}
\eea

If we choose the initial monopole to be the unstable configuration
having $q^a = q^a_i = (0,4/g)$, then $q^2_i = 16\gamma^2
/ g^2$. It is for this configuration that we tune the 6D
cosmological constant to obtain a flat 4D spacetime, $\lambda =0$.
This fixes the initial radius of the internal sphere and the 6D
cosmological constant to be
\be\label{Sol1}
 a^2_i = \frac{\kappa^2 q_i^2}{8} = \frac{2 \,
 \kappa^2 \gamma^2}{g^2}
 \qquad \hbox{and} \qquad
 \Lambda = \frac{\kappa^2 q_i^2}{8 a_i^4} = \frac{1}{a_i^2}
 = \frac{g^2}{2\, \kappa^2 \gamma^2}\,.
\ee

To find the endpoint, we now take the monopole charge to be the
topologically stable one having $N=-1$ and so $q^a_f = (0,-2/
 g)$, and so $q^2_f = 4\gamma^2/g^2 = \frac14 \,
q^2_i$. Since $\Lambda$ is no longer free to be adjusted, we now
solve eqs.~(\ref{condi1}) and (\ref{condi2}) for the final radius,
$a_f$, of the 2D sphere, and the curvature, $\lambda_f$, of the
final 4D spacetime.

We find in this way that the 4D curvature becomes
\be
 3 \lambda_f = \frac{1}{4 a_i^2}
 - \frac{\kappa^2 q_f^2}{32\,a_f^4}
 = \frac{1}{4 a_i^2}
 - \frac{a_i^2}{16 \, a_f^4} \,,
\ee
while the new radius of the 2-sphere is given by
\be
 \frac{1}{a_f^2} = \frac{1}{4 a_i^2}
 + \frac{3\,\kappa^2 q_f^2}{32\,a_f^4}
  = \frac{1}{4 a_i^2}
 + \frac{3\, a_i^2}{16 \, a_f^4} \,.
\ee
This has two roots, given by
\be\label{radii}
   a_{f\pm}^2 = 2\,a_i^2 \left[ 1\pm \sqrt{1-\frac{3}{4}
         \frac{q_f^2}{q_i^2}}  \right] = 2\,a_i^2
    \left[ 1\pm \frac{1}{4}\sqrt{13}  \right] \,,
\ee
and so $a_{f+} \simeq 1.95 \, a_i$ while $a_{f-} \simeq 0.444 \,
a_i$. The corresponding 4D curvature then becomes
\be
 3\lambda_{f\pm} = \frac{1}{a_i^2} \left( \frac{7 \pm 2
 \sqrt{13}}{29 \pm 8 \sqrt{13}}\right) \,,
\ee
and so $\lambda_{f+} \simeq 0.0819/a_i^2$ and $\lambda_{f-} \simeq
-0.452/a_i^2$. Clearly the radius of the sphere increases for the
dS solution and decreases for the AdS one.

\subsubsection*{Energetics}

Intuitively, one would expect the AdS case to be the natural
endpoint, since one expects to obtain a negative potential energy
after lowering it below the initially zero value needed to ensure
a flat 4 dimension, as we next check explicitly. To this end
define the potential energy (per unit 3D volume), $\mathcal{E}$,
of the effective 4D theory as the sum of the static 6D energy
({\it i.e.} gradient, magnetic and potential energy), integrated
over the extra dimensions, with
\be\label{E1}
 {\mathcal E} = \frac{1}{\kappa^2}\int{\exd^2x \sqrt{g_2} \;
 \left[ -R_{(2)}
 + \frac{\kappa^2}{4} \, \Tr \, F_{mn} F^{mn} + \Lambda \right]} \,.
\ee
Here $R_{(2)} = {2}/{a^2}$ denotes the 2D curvature scalar, while
the magnetic energy  goes as
$\Tr \,F_{mn} F^{mn} = {q^2}/{2a^4}$,
leading to
\bea\label{energy1}
 {\mathcal E} &=& \frac{4\pi a^2}{\kappa^2} \left[  -\frac{2}{a^2} +
 \Lambda + \frac{\kappa^2q^2}{8\,a^4} \right] \nonumber\\
 &=& \left( \frac{4 \pi a^2}{\kappa^2} \right)^2
 \frac{\kappa^2}{4\pi} \left[ - \frac{2}{a^4}
 + \frac{\Lambda}{a^2} + \frac{\kappa^2 q^2}{8 \, a^6}
 \right] \,.
\eea

The second equality of eq.~(\ref{energy1}) pulls out four powers
of the 4D Planck mass, $M_p^2 = 4\pi a^2/\kappa^2$, which is
useful when verifying that $\mathcal{E}$ as defined reproduces the
correct equations of motion when used in the 4D field theory. It
is useful to display the factors of $M_p$ explicitly in this way
because transforming to the 4D Einstein frame ensures that these
are held fixed when the 4D field $a$ is varied to minimize the
potential energy, leading to the condition
\be \label{4Daeq}
 \frac{\partial (\mathcal{E}/M_p^4)}{\partial a}
 = \frac{2\kappa^2 }{\pi a^3} \left[ \frac{1}{a^2}
 - \frac{\Lambda}{4}
 - \frac{3 \kappa^2 q^2}{32\, a^4} \right]
 = 0 \,,
\ee
in agreement with eq.~(\ref{condi2}) determining $a$. The 4D
Einstein equations similarly equate $R_{\mu\nu} = 3 \lambda
g_{\mu\nu}$ to $\left( \mathcal{E}/M_p^2 \right) g_{\mu\nu}$,
leading to the condition
\bea
 6 \lambda &=& - \frac{2}{a^2} + \Lambda
 + \frac{\kappa^2 q^2}{8\, a^4} \nonumber\\
 &=& \frac{\Lambda}{2} - \frac{\kappa^2 q^2}{
 16\, a^4} \,,
\eea
where the last equality --- which agrees with eq.~(\ref{condi1})
--- uses the field equation, eq.~(\ref{4Daeq}), for $a$.

We may now compare the value of $\mathcal{E}$ when evaluated at
the initial and final configurations considered above. Evaluating
using our previous results for $a_i$, $q_i^2$ and $\Lambda$ leads
to $\mathcal{E}_i = 0$ for the initial unstable solution,
consistent with having tuned $\Lambda$ to ensure the flatness of
the initial 4D geometry. For the two candidate endpoint solutions,
the potential energy is instead
\bea
 {\mathcal E}_{f\pm} &=& \frac{4\pi}{\kappa^2} \left[
 -2 + \Lambda a_{f\pm}^2 + \frac{\kappa^2 q_f^2}{
 8\, a_{f\pm}^2} \right] \nonumber\\
 &=& \frac{4\pi}{\kappa^2} \left[ -2 + \left(
 \frac{a_{f\pm}}{a_i} \right)^2 + \frac{q_f^2}{q_i^2}
 \left( \frac{a_i}{a_{f\pm}} \right)^2 \right] \\
 &=& \frac{2\pi}{\kappa^2} \left( \frac{14 \pm 4 \sqrt{13}}{
 4 \pm \sqrt{13}} \right) \,,
\eea
and so $\mathcal{E}_{f+} \simeq 3.74 \left( 2\pi/\kappa^2 \right)
> 0$ and $\mathcal{E}_{f-} \simeq -1.07 \left( 2\pi/\kappa^2
\right) < 0$. Clearly we have $a_{f+} > a_i > a_{f-}$, and
${\mathcal E}_{f+} > {\mathcal E}_i >  {\mathcal E}_{f-}$,
indicating that the endpoint solution reached after the
instability indeed corresponds to a shrunken extra-dimensional
sphere together with 4D AdS space.


\section{Endpoint Revisited: Including the Dilaton}

We next reconsider the problem of direct interest for higher
dimensional supergravity, by supplementing the Einstein-YM theory
with an appropriate scalar dilaton field. In this case we find the
endpoint configurations do not preserve the maximal symmetry of
the underlying 4D and/or 2D geometries of the original unstable
monopole-supported system. As emphasised in \cite{GibbonsMaeda},
the presence of the dilaton crucially changes the dynamics of the
system, and this considerably complicates the search for the new
endpoint solutions.

To see why the dilaton is so different we again start with a
monopole-supported solution to 6D chiral gauged supergravity with
couplings chosen to allow a $\hbox{Mink}_4 \times S_2$ solution
with 4 flat large dimensions, see Eq. (\ref{E:sphere}-\ref{rugbymonopole}). Now, however, if
the monopole decays 
to its stable topological cousin, a new maximally symmetric
solution supported by the stable monopole no longer satisfies the
field equations (\ref{EOM}), which require $\Box \sigma = 0$ together
with the Einstein and
Yang-Mills equations. In detail, the equations of motion under the
maximally symmetric ansatz $ds^2 = \exd s_4^2 + a^2 (\exd \theta^2 +
\sin^2\theta \exd\phi^2)$, $F_{\theta\phi} = \frac{q^a \,
  Q_a}{2}\sin\theta$ and $\sigma = \sigma_0 = const$, with $\exd s_4^2$
the metric on 4D (A)dS or Minkowski spacetime, 
together imply the 4D curvature $\lambda=0$, $a \, e^{-\kappa\sigma_0/4} =
\kappa/2\sqrt{2}g_1$ and $q^2 = 1/g_1^2$.  This is to be compared with
the initial configuration Eq. (\ref{E:sphere}-\ref{rugbymonopole}).  So, although one 
combination of the parameters, say $a \, e^{\kappa\sigma_0/4}$, is left
free thanks to the classical scaling symmetry, the magnetic flux in particular is fixed to its original --
unstable -- magnitude if we insist on keeping the maximal symmetries.
This shows that once the monopole flux decays the dilaton gradient, $\partial_\M \sigma$, is
necessarily nonzero, thereby picking out preferred directions in
the underlying spacetime.

A key question asks whether this gradient points in the compact
two directions, $\partial_m \sigma \ne 0$, or in the large
spacetime directions, $\partial_\mu \sigma$. In this section we
first argue that the system is likely to prefer growing nonzero
gradients in the large 4 dimensions, and then describe the
relative merits of two classes of candidate endpoint solutions
that break the 4D spacetime symmetries: a one-parameter family of
supersymmetric solutions \cite{glps}; and a class of new solutions
to which one is led by adapting the arguments of the previous
section to include the dilaton.

\subsection{4D or 2D: Which symmetries break?}

We now argue that for 6D supergravity it is the 4D spacetime
symmetries that generically prefer to break. If true this is
somewhat surprising, since the instability revealed by the
linearised analysis is in modes that vary in the internal 2
dimensions and not the macroscopic 4 dimensions. However, it is
known \cite{bulksings} that all of the axially symmetric bulk
solutions having AdS 4D geometry necessarily have a curvature
singularity in the 2D geometry at the position of one of the two
source branes.\footnote{This is also a corollary of the fact
\cite{bulksings} that all of the solutions having only conical
singularities at the branes have 4D geometries that are flat.} Any
decay to a solution of the form AdS${}_4 \times M_2$ therefore
necessarily requires the development of a curvature singularity in
the 2D geometry at the position of one of the source branes, even
if the initial unstable solution has no such a singularity. But
the divergence of bulk fields at a singularity is related to the
physical properties of the brane which is situated there
\cite{UVCaps,BdRHT}, with a curvature singularity in particular
implying a brane coupling to the bulk dilaton. Since it is not
clear how such a change to intrinsic brane properties can be
triggered by the decay of a monopole in the bulk, we instead
explore the possibility that it is the 4D spacetime symmetries
that break.

The simplest way to see the necessity for a curvature singularity
is to recognize that the effective 4D potential energy turns out
to depend only on the near-brane limit of the $\sigma$ field when
its derivatives, $\partial_m \sigma$, point purely along the 2D
directions \cite{ocho}. That is, we evaluate
\be
 \mathcal{E} = \int \exd^2 x \sqrt{g_2} \; \left[
 - \frac{1}{\kappa^2} \, R_{(2)} + \frac14 \, \partial_m \sigma \,
 \partial^m \sigma + \frac14 \, e^{\kappa \sigma/2} \, \Tr
 F_{mn} F^{mn} + \frac{8 g_1^2}{\kappa^4} \, e^{-\kappa \sigma/2}
 \right] \,, \label{4DV}
\ee
at an arbitrary solution to the field equations, eqs.~(\ref{EOM}),
assuming only that all tensor components point purely along the
compact 2 dimensions. Use of the Einstein and dilaton equations in
particular then show \cite{ocho} that
\be \label{Eastotalderiv}
 \mathcal{E} = - \frac{1}{2\kappa} \int \exd^2 x \sqrt{g_2} \;
 \Box \sigma \,,
\ee
which vanishes on a smooth manifold. For example, when evaluated
for the particular solutions of eqs.~(\ref{E:8author}) and
(\ref{E:8author1}), we find (keeping in mind the conical
singularities at $\rho = \rho_\pm$)
\be
 \mathcal{E} = - \frac{\pi}{\kappa} \int_{\rho_-}^{\rho_+}
 \exd \rho \; \partial_\rho \Bigl[ \sqrt{g_2} \; \partial^\rho
 \sigma \Bigr]
 = \frac{2\pi}{\kappa^2} \left[ \frac{h(\rho_-)}{\rho_-}
 - \frac{h(\rho_+)}{\rho_+} \right] = 0 \,.
\ee

In the presence of singularities localized at source branes, each
brane can be isolated within a small circle that acts as the
boundary of the bulk geometry, leading the right-hand-side of
eq.~(\ref{Eastotalderiv}) to evaluate to a sum of terms involving
the radial dilaton derivative, $n \cdot \partial \sigma$,
evaluated at the brane positions. But the presence of such a
nonzero scalar gradient near the codimension-2 brane requires
$\phi$ to diverge logarithmically there, and the stress energy of
this configuration makes the curvature also diverge. This argument
is in agreement with the explicitly known solutions of
ref.~\cite{bulksings}.

But the near-brane dilaton derivative is related
\cite{UVCaps,BdRHT} by the bulk-brane matching conditions to the
effective codimension-2 brane tension, $T_2(\phi)$, with $n \cdot
\phi$ being proportional to its derivative $T_2'$. As such, the
near-brane dilaton derivative cannot change without there also
being a physical change to the source branes, making such a
configuration an unlikely endpoint for an unstable monopole. 
This being said, we shall also find the necessity of new types of
singularities in some solutions breaking the 4D symmetries, and so this
argument cannot be regarded as decisive until the interpretation of those
singularities is better understood.

With
this motivation we next examine two categories of candidate
endpoint solutions that break the 4D symmetries.

\subsection{Supersymmetric $\hbox{AdS}_3 \times \tilde S_3$}

Supersymmetric solutions are always attractive options when
seeking stable endpoints from initially unstable initial
configurations, and it is the remarkable scarcity of such
solutions having the form $M_4 \times M_2$, with $M_4 =
\hbox{Mink}_4$ or $\hbox{AdS}_4$, that helps make the endpoint of
monopole decay in 6D supergravity such a puzzle. The only known
solutions of this type have $M_4 = \hbox{Mink}_4$, $M_2 = S_2$,
and align the monopole in the $U(1)_R$ directions with monopole
number $N=\pm 1$ \cite{SS}.

Other supersymmetric solutions do exist \cite{glps}, however, they
just have fewer 4D spacetime symmetries. These solutions have
geometries $\hbox{AdS}_3 \times \tilde S_3$, where $\tilde S_3$
denotes a one-parameter family of `squashed' 3-spheres. The field
configurations have constant dilaton, $\partial_\M \sigma = 0$,
and
\bea
 \exd s^2 &=& \exd s^2_{\rm AdS{}_3} + a^2 \left( \omega_1^2
 + \omega_2^2 \right) + b^2 \omega_3^2 \nonumber\\
 H_3 &=& \xi \left( \omega_1 \wedge \omega_2 \wedge \omega_3
 + \frac{\varepsilon_3}{a^2 b} \right)  \\
  F_2 &=& k \, \omega_1 \wedge \omega_2 \nonumber\,,
\eea
where $\exd s^2_{\rm AdS{}_3}$ is the line-element for
$\hbox{AdS}_3$ and $\varepsilon_3$ denotes the volume 3-form for
the internal 3D geometry. The $\omega_m$ denote the left-invariant
1-forms on the 3-sphere, that satisfy $\exd \omega_m = - \frac12
\, \epsilon_{mnp} \, \omega_n \wedge \omega_p$, and so
\be
 \omega_1 + i \omega_2 = e^{-i \psi} \left( \exd\theta
 + i \sin\theta \, \exd \phi \right) \,, \qquad
 \omega_3 = \exd \psi + \cos \theta \, \exd \phi \,,
\ee
where $(\theta, \phi, \psi)$ are Euler angles on the 3-sphere.

The equations of motion impose the following three relations among
the solution's four parameters, $a$, $b$, $\xi$ and $k$
\cite{glps}:
\be
 b^2 = \kappa^{5/2} \xi\,, \qquad
 a^2 = \frac{\kappa^3 k}{4g_1} = \frac{1}{16 g_1^2}
 \left( 1 \pm \sqrt{1 - 32 \kappa^{1/2} g_1^2 \xi}
 \right) \,,
\ee
in terms of which the AdS${}_3$ Ricci tensor is $R_{\alpha \beta}
= 2 \lambda \, g_{\alpha \beta}$ with $\lambda = -b^2/(4a^4)$.

Is this the endpoint of the evolution away from the unstable
monopole? Such a scenario would be very attractive, indicating a
dynamic spontaneous compactification wherein the monopole
instability triggers one of the large 4 dimensions to roll up into
one of the directions in $\tilde S_3$. And because $\lambda$ is
negative this might be argued to be favoured energetically in
terms of an appropriate 3D potential energy. Better yet, the
supersymmetric $\hbox{Mink}_4 \times S_2$ solution can be obtained
formally from the $\hbox{AdS}_3 \times \tilde S_3$ solutions by
taking the limit $b \to 0$ \cite{glps}, indicating there might be
a plausible path through field space leading from the initial
unstable configuration to the final supersymmetric one.

There are a number of possible objections to the proposal that
these solutions represent to endpoint of the monopoles of present
interest, however. Not least, the natural way to obtain 4 large
directions from AdS${}_3 \times \tilde S_3$ is by taking the lone
squashed direction to become large, $b \gg a$, rather than taking
$b \ll a$. However in the limit $b \gg a$ the curvature of the
large 3 dimensions becomes larger and not smaller, and there is
furthermore an obstruction to taking this limit within the
supersymmetric solutions since it formally would require taking
$\xi$ very large, but $a^2$ becomes complex in this limit. We
therefore next seek other options for the decay endpoint.

\subsection{Deking  the Dilaton}
\footnote{{\bf deke} {\it v.} (in ice hockey) to draw a defending
player out of position by faking a shot or movement: {\it deked
the goalie with a move from left to right.}}
As noted above, it is the dilaton that appears to prevent the
system's relaxation towards a maximally symmetric solution, and so
removes the attractive picture obtained in the EYM system
described in \S3. In the remainder of this section we use an
elegant trick \cite{tanaka} that reformulates the EYM-dilaton
system as a dilaton-free system in higher dimensions. We do so
with the goal of exploring whether the analysis of \S3\ can lead
to a better candidate endpoint, for which the preserved maximal
symmetries involve the fictitious dimensions associated with the
dilaton rather than being part of the physical 6 dimensions of our
starting supergravity.

\begin{figure}[htbp]
\begin{center}
\includegraphics[scale=0.45]{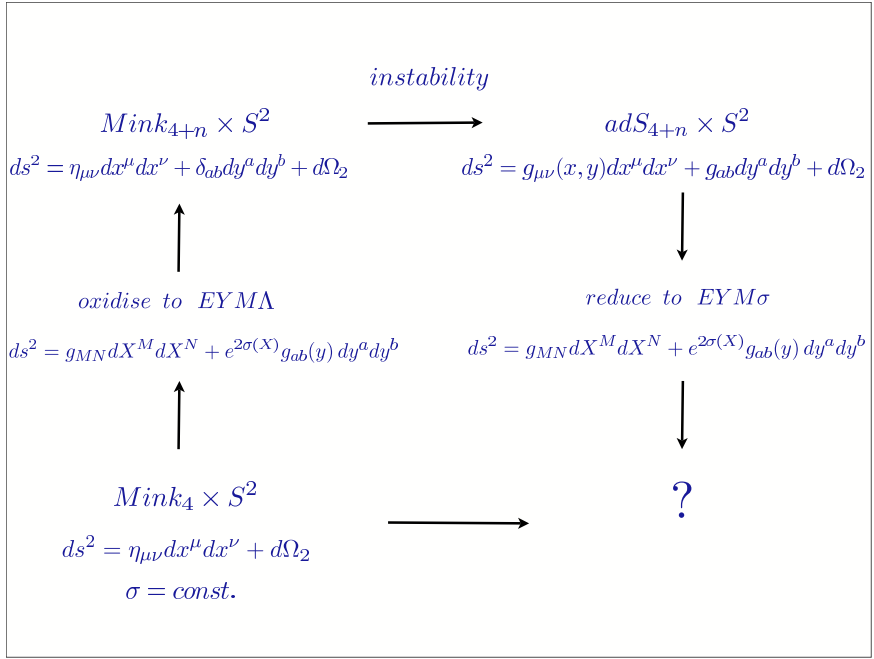}
\caption{The oxidation-reduction cycle used to generate solutions
in the 6D dilatonic theory.} \label{fig:oxdred}
\end{center}
\end{figure}

The idea behind the trick is that the dilaton can be regarded as a
modulus obtained by compactifying a simpler system in higher
dimension.\footnote{A similar logic underlies the discussion of
$F$-theory vacua in Type IIB string compactifications having
nontrivial dilaton profiles.} In particular, we consider EYM
theory in $(6+n)$ dimensions, chosen so that its dimensional
reduction to 6D leads to the dilaton-EYM theory of interest. With
care, solutions in the higher dimensional non-dilatonic theory can
be reduced to solutions of dilatonic Einstein Yang-Mills in six
dimensions, and 6D supergravity configurations can be `oxidised'
to higher dimensional solutions. Ref. \cite{Masato} performs a
similar analysis to study the dynamics of instabilities in warped
de Sitter solutions to 6D dilatonic Einstein Maxwell
theory, building on studies of the dilaton-free model \cite{Chethan}.

This trick is useful because the stability analysis of previous
sections can be translated word-for-word to the higher dimensional
system, at least for unwarped backgrounds.\footnote{The bilinear
action for the relevant modes is identical in this case (see
equation (44) of \cite{seif}).} In particular, an unstable
monopole-supported configuration with geometry $\hbox{
Mink}_{(4+n)} \times S_2$, is unstable for large enough magnetic
quantum numbers, and applying the arguments of \S3\ to the
higher-dimensional system indicates a decay to $\hbox{AdS}_{(4+n)}
\times S_2$ supported by a stable monopole. The logic (illustrated
in Fig.~\ref{fig:oxdred}) then is to dimensionally reduce both the
unstable solution and its stable endpoint down to 6D to find the
corresponding transition to which this points in the
lower-dimensional dilaton system.

\subsubsection{Oxidation/Reduction}

To proceed in detail we start with the $D=6+n$ action
\cite{tanaka, Masato}
\be\label{Daction}
 S= \frac{1}{\kappa_\D^2} \int \exd^\D x \sqrt{-g_\D}
 \left[ R_{(\D)} -\frac{\kappa_\D^2}{4} \, \Tr \, {\mathcal F}^2
 - \Lambda \right] \,,
\ee
whose equations of motion are
\bea
 && R_{\Lm\Gm} = \frac{\kappa^2_\D}{2} \, \Tr \left(
 {\mathcal F}_{\Lm\Om} \mathcal{F}_\Gm^{\,\,\,\Om} \right)
 + \frac{g_{\Lm\Gm}}{(D-2)} \left(
    \Lambda - \frac{\kappa^2_\D}{4}
    \,\Tr \, {\mathcal F}^2 \right)  \\
 && \nabla_\Lm {\mathcal F}^{\Lm\Gm} - i {g}
   \mathcal{A}_\Lm {\mathcal F}^{\Lm\Gm} = 0 \,. \nonumber
\eea

To dimensionally reduce we seek solutions to these equations
having the form\footnote{The indices $\Gamma,\Lambda,..$ run from
0 to $n+5$, while indices $a,b,..$ run from 1 to $n$ and 6D
indices $M,N,..$ run from 0 to 5 as before. We reserve $A,B,..$ to
run from 0 to $n+3$ in later applications. }
\bea
 \exd s_\D^2 = g_{\Lm\Gm} \, \exd x^\Lm \exd x^\Gm
 &=& \hat g_{\M\N}(x) \, \exd x^\M \exd x^\N
 + e^{2 \varphi(x)} g_{ab}(y) \, \exd y^a \exd y^b \nonumber \\
 &=& e^{-n \,\varphi(x)/2} g_{\M\N}(x)
 \, \exd x^\M \exd x^\N + e^{2\varphi(x)} g_{ab}(y)
 \, \exd y^a \exd y^b  \\
 {\mathcal F}_{\M\N} = {\mathcal F}_{\M\N}(x)
 \quad &\hbox{and}& \quad {\mathcal F}_{a\M} =
 \mathcal{F}_{ab} = 0 \,. \nonumber \label{truncansatz}
\eea
Here $g_{ab}(y)$ is an $n$-dimensional maximally-symmetric metric,
whose curvature scalar is: $g^{ab} R_{ab} = n(n-1)K$,
for constant $K$.
Furthermore, the above configuration is the most general one
consistent with this maximal symmetry, which ensures that
solutions to the truncated action are guaranteed also to be
solutions of the full higher-dimensional equations. (Such a
configuration is called a `consistent' truncation \cite{KKK}.)

With this ansatz the action of the truncated 6D theory becomes
\cite{tanaka, Masato}
\bea
 S &=& \frac{1}{\kappa^2} \int \exd^6 x \sqrt{-g} \;
 \left[  R - \frac{n(n+4)}{4} \, \partial_\M \varphi
  \,\partial^\M \varphi - \frac{\kappa^2}{4} \,
  e^{n\varphi/2} \, \Tr \, F^2  \right.\\
  &&\qquad\qquad\qquad \qquad\qquad\qquad
  \qquad  \left. \phantom{\frac12} - \Lambda
  e^{-n\varphi/2} + K n(n-1) \, e^{-(n+4)
  \varphi/2} \right]  \,, \nonumber
\eea
where we define $\kappa^2 := \kappa_D^2/V$ and $F_{\M\N} :=
V^{1/2} {\mathcal F}_{\M\N}$, with $V$ the volume of the
$n$-dimensional manifold computed with the metric $g_{ab}$.
Finally, defining
\be\label{DSol2}
 \kappa \,\sigma = \sqrt{n(n+4)} \; \varphi \,
 \qquad \hbox{and} \qquad
 \zeta^2 = \frac{n}{n+4} \,,
\ee
and so $n = 4 \zeta^2/(1-\zeta^2)$, allows the action to be
written
\bea
 S &=& \frac{1}{\kappa^2} \int \exd^6 x \sqrt{-g} \;
 \left[ R - \frac{\kappa^2}{4} \, \partial_\M \sigma
 \, \partial^\M \sigma - \frac{\kappa^2}{4} \,
 e^{\zeta \,\kappa\sigma/2} \, \Tr \, F^2  \right.\\
  &&\qquad\qquad\qquad \qquad\qquad\qquad
  \qquad  \left. \phantom{\frac12}
 - \Lambda \, e^{-\zeta\,\kappa\sigma/2}
 + K \, \frac{4\zeta^2(5\zeta^2-1)}{1-\zeta^2} \,
 e^{-\kappa\,\sigma/2\zeta}\right]  \,. \nonumber
\eea
This shows that the 6D supergravity action, eq.~(\ref{SB}), is
obtained in the formal limit where $K = 0$ and $\zeta \rightarrow
1$ (and so $n \to \infty$), provided we also identify $\Lambda =
8g_1^2/\kappa^2$.

\subsubsection{The Rugby Ball and its Decay}

As an application consider the following simple monopole-supported
compactification from $D$ to $D-2$ dimensions:
\bea
 \exd s^2_\D &=& g_{\A\B}\, \exd x^\A \exd x^\B
 + a^2 (\exd\theta^2 + \sin^2{\theta} \, \exd\phi^2) \nonumber\\
  {\mathcal F}_{\theta\phi}^a &=& \frac{q^a_\D}{2} \, \sin{\theta}
\eea
where $A,B,.. = 0,1,...,n+3$, for which directions the $d = (D-2)
= (4+n)$-dimensional metric is
\be
 R_{\A\B} = (d-1) \lambda_d \, g_{\A\B}
 = (D-3) \lambda_d \, g_{\A\B} \,.
\ee

Using this ansatz in the $D$-dimensional equations of motion
allows their content to be boiled down to
\bea\label{DSol22}
 (D-3) \lambda_d &=& -\frac{1}{D-2}
 \left[ \frac{\kappa^2_\D q_\D^2}{8 a^4} -\Lambda \right]  \\
 \frac{1}{a^2} &=& \frac{\kappa^2_\D q_\D^2}{8a^4}
 + (D-3) \lambda_d \,,
 \label{DSol1}
\eea
whose solutions are
\bea \label{radios1}
 a^2_\pm &=& \frac{(D-2)}{2\Lambda} \left[ 1\pm
 \sqrt{1- \frac{(D-3)}{2(D-2)^2} \, \Lambda \, \kappa^2_\D q_\D^2}
 \right] \\
 \lambda_d &=& \frac{1}{(D-3)^2} \left[ \Lambda
 - \frac{1}{a^2_{\pm}} \right] \,. \label{lambda1}
\eea
Eliminating $q_\D^2$ gives the 2-sphere radius in terms of the
$d$-dimensional curvature:
\be\label{radios2}
 a^2_\pm = \frac{1}{\Lambda - (D-3)^2\lambda_d}  =
         \frac{1}{\Lambda - (n+3)^2\lambda_d} \,.
\ee

Applying these results to an initial geometry $\hbox{Mink}_4
\times S_2$ supported by an unstable monopole having charge $q^2 =
q^2_i$ shows that the parameter $\Lambda$ must be tuned to
\be\label{azero}
 \Lambda = \frac{1}{a_i^2} = \frac{8}{\kappa^2_\D q_{\D i}^2}
 = \frac{8}{\kappa^2 q_i^2} \,,
\ee
in which the final equality cancels the factors of
extra-dimensional volume, $V$, that appear in the relations
between the $D$- and 6-dimensional versions of $\kappa$ and $q^2$.
Dimensionally reducing this geometry on $n$ of the flat directions
then trivially reproduces the rugby-ball solution,
eq.~(\ref{E:sphere}), of 6D supergravity (whose decay we wish to
study).

As in \S3 we suppose the endpoint of the instability in the
$D$-dimensional system also to be given by solutions to these same
equations, but for the smaller charge of the final stable
monopole:
$q_{\D\, f}^2 < q_{\D\,i}^2$. And eq.~(\ref{radios1}) implies that
shrinking $q^2$ makes $a_+^2$ get larger while $a_-^2$ gets
smaller, which eq.~(\ref{lambda1}) in turn implies $\lambda_{d+}$
is positive while $\lambda_{d-}$ is negative. As in \S3\ this
predicts the endpoint to be a smaller monopole-supported sphere,
with negatively curved large directions.

The idea now is to dimensionally truncate the endpoint
monopole-supported $D$-dimensional geometry on $n$ of its AdS
dimensions, thereby obtaining a candidate endpoint solution for
the 6D EYM-dilaton system. To this end it is useful to rewrite the
$D$-dimensional metric in terms of flat spatial slicings
\bea
 \exd s^2_\D &=& \Bigl[ \exd x^2 + e^{2 \sqrt{-\lambda_d} \,x}
 \left( -  \exd t^2 + \delta_{ij} \,
 \exd x^i \exd x^j + \delta_{ab} \, \exd y^a \exd y^b
 \right) \Bigr]
 + a^2_- \, \exd\Omega^2_2  \nonumber\\
 &=& e^{-n\sqrt{-\lambda_d} \,x/2}
 \left[ - e^{(2+n/2) \sqrt{-\lambda_d} \,x} \, \exd t^2
 + e^{n\sqrt{-\lambda_d} \,x/2} \, \exd x^2
 +  e^{(2+n/2)\sqrt{-\lambda_d} \,x} \, \delta_{ij} \,
 \exd x^i \exd x^j \right.  \nonumber \\
 && \left. \hskip5cm+ a^2_- e^{n\sqrt{-\lambda_d} \,x/2} \,
 \exd\Omega^2_2 \right] + e^{2 \sqrt{-\lambda_d} \,x}
 \, \delta_{ab} \, \exd y^a \exd y^b \,,
\eea
where $i,j,..$ run from 1 to 2, while (as before) $a,b,..$ run
from 4 to $4+n$, and $\exd \Omega^2_2$ denotes the standard metric
on the unit 2-sphere.

Comparing this last expression with the ansatz,
eq.~(\ref{truncansatz}) allows the dilaton to be read off from the
$x$-dependence of the $n$-dimensional truncated metric, giving
$\varphi = \sqrt{-\lambda_d}\, x$, or
\be
 \kappa \,\sigma = \frac{4\zeta}{1-\zeta^2} \sqrt{-\lambda_d} \, x \,.
\ee
Using this in eq.~(\ref{truncansatz}) then also allows the 6D
metric to be identified. Making the change of variables
\be
 z = \int{e^{n\sqrt{-\lambda_d}\,x/4} \exd x }
 = \frac{4}{n\sqrt{-\lambda_d}} \,
 e^{n\sqrt{-\lambda_d}\,x/4} \label{z}
\ee
allows the truncated 6D solution to be written
\be
 \exd s^2 = -\left( \frac{z}{L_n} \right)^{2+8/n}
 \, \exd t^2 + \exd z^2
 + \left( \frac{z}{L_n} \right)^{2+8/n}
 \, \delta_{ij} \, \exd x^i \exd x^j
 + \left( \frac{z}{L_n} \right)^{2}
 \, a^2_-  d\Omega^2_2  \,,
\ee
and
\be
 \kappa \, \sigma = \frac{4}{\zeta} \ln{ \left(
 \frac{z}{L_n} \right)} \,,
\ee
where the length scale $L_n$ is defined by
\be
 \frac{1}{L_n} := \frac{n\sqrt{-\lambda_d}}{4}
 = \frac{n}{4(n+3)} \sqrt{\frac{1}{a^2_-}
 - \frac{1}{a_i^2}} \,, \label{b} \ee
and the expression for $\lambda_d$ in terms of $a_-$ and $a_i$ is
used. The final step is to take $n \to \infty$ ($\zeta \to 1$) to
recover 6D Nishino-Sezgin supergravity. Both $L$ and $a_-$ remain
finite in this limit, with
\be
 \frac{1}{L} = \lim_{n \to \infty} \frac{1}{L_n}
 = \frac{1}{4} \sqrt{\frac{1}{a_-^2}
 - \frac{1}{a_i^2}}
\ee
and
\be
 \lim_{n \to \infty} a_-^2 = \lim_{n \to \infty}
 \frac{(n+4)}{2\Lambda} \left[ 1 \pm \sqrt{1-
 \frac{(n+3)}{2(n+4)^2} \, \Lambda \, \kappa^2 q_f^2}
 \right]
 = \frac{ \kappa^2 q_f^2}{8} \,.
\ee

Combining all expression gives the final result for the candidate
endpoint solution to gauged chiral 6D supergravity
\bea\label{endsolution}
 && \exd s^2 = \exd z^2 + \left( \frac{z}{L} \right)^2 \,
 \Bigl[ -\exd t^2
  + \delta_{ij} \, \exd x^i \exd x^j \Bigr]
 + \left( \frac{z}{L} \right)^2 \, a^2_-
 \exd\Omega^2_2 \,, \nonumber\\\nonumber \\
 && \kappa \, \sigma = 4 \ln{ \left( \frac{z}{L} \right)}
 \qquad \hbox{and} \qquad F_{\theta\phi}^a =
 \frac{q^a_f}{2} \,\sin{\theta} \,.
\eea
One can check directly that this configuration indeed solves the
6D supergravity equations, and in fact can be recognized as one of
the scaling solutions found in \cite{scaling}, but with the
scaling occurring along a 4D spatial coordinate, $z$, rather than
time. Also noteworthy is the relation this solution implies
between the sphere's radius, $r$, and the dilaton: $r^2 =
e^{\kappa \sigma/2} a_-^2$, which is also familiar (but
$z$-independent) from the Salam-Sezgin compactification \cite{SS}.

The solution eventually breaks down for small $z$ due to the
singularity as $z \to 0$, where both the dilaton and the 6D Ricci
scalar,
\be
 R = \frac{2(L^2 - 10 \, a_-^2)}{z^2 a_-^2} \,,
\ee
blow up. Since this singularity has no counterpart in the higher
dimensional $\hbox{AdS}_{4+n} \times S_2$ EYM solution, its
emergence is a consequence of taking the limit $n \rightarrow
\infty$. The finite-$n$ geometries may be regarded in this way as
providing resolutions of this singularity, along the lines of the
higher-dimensional resolution of dilatonic black hole
singularities in string theory described in
ref.~\cite{dilatonicbh}.

At large $z$ the radius of the compact 2-sphere becomes very
large, implying an eventual breakdown of the 4D effective theory
even at very low energies. It is instructive to ask how the metric
varies in the 4D Einstein frame, especially since the dependence
on $z$ only arises as an overall conformal factor (as may be seen
using the coordinate change $u = \ln{(z/L)}$),
\be
 \exd s^2 =  e^{2u} \Bigl( \eta_{\mu\nu} \,
 \exd x^\mu \exd x^\nu + a^2_- \, \exd \Omega^2_2 \Bigr) \,,
\ee
implying the breaking of the 4D maximal symmetry therefore drops
out of conformally invariant quantities.
Since the volume of the 2
compact dimensions varies as $V_2 = (z/L)^2 a_-^2 \propto e^{2u}$,
the 4D Einstein frame metric scales with $u$ as $g_{\mu\nu}^{(E)}
= e^{2u} g_{\mu\nu}$, which is again $u$-dependent, and in fact turns
out to be the same geometry as that of the 6D Einstein frame.

\subsubsection*{Stability}

The stability of this solution follows from that of the
corresponding oxidised solution, $\hbox{AdS}_{4+n} \times S_2$,
since the fluctuations in the 6D model are a sub-sector of those
in the oxidised model, allowing us to conclude that our proposed
endpoint is a stable solution, without performing the linearised
stability analysis from scratch. Fluctuations in the $(6+n)$D EYM
model divide into two decoupled sectors:
\begin{enumerate}
\item The metric fluctuations and gauge field fluctuations in the
direction of the $U(1)$ monopole in the Lie Algebra.  These were
studied in \cite{bousso}, where they were found to be stable, in
the sense that none violate the Breitenlohner-Freedman bound.
\item The gauge field fluctuations orthogonal to the $U(1)$
monopole. We argued above that the presence of instabilities in
this sector for $\hbox{Mink}_4 \times S_2$
\cite{seifinstabilities, prs2} generalise to higher dimensions and
so these modes are also unstable in the $\hbox{Mink}_{4+n} \times
S_2$ theory. The identical argument shows that stable monopoles in
$\hbox{Mink}_{4} \times S_2$ oxidise to configurations that are
also stable in $\hbox{Mink}_{4+n} \times S_2$. The same conclusion
should also apply for $\hbox{AdS}_{4+n} \times S_2$, since the
Kaluza-Klein mass operator does not depend on the curvature of the
external geometry, but only on the curvature of the internal
geometry and the internal flux.
\end{enumerate}

\subsubsection*{Energy}

The higher-dimensional picture also argues for there being an
energetic criterion which favours these new solutions as having
smaller energy then the initial, unstable one.
Given the non-trivial profile for the dilaton in the large dimensions,
an appropriate definition for the energy is the sum of the 4D dilatonic
gradient energy and the potential energy of the 4D effective theory
defined in (\ref{4DV}),
which emerges from the gradient, magnetic and potential energy in the
extra dimensions.   Integrating out the extra dimensions, the
total energy density (per unit 3D volume) in the Einstein frame,
$g^{(E)}_{\mu\nu} =  e^{\kappa\sigma/2} \, g_{\mu\nu}$, is:
\be
{\mathcal E}_{TOT} = \frac{e^{-\kappa\sigma}}{\kappa^2} \int \exd^2 x
\, \sqrt{g_2} \;
\left[ \frac{\kappa^2}{4} \, e^{\kappa\sigma/2}\, \partial_z \sigma \,
  \partial^z \sigma - R_{(2)}
  + \frac{\kappa^2}{4} \Tr \, F_{mn}\, F^{mn} +  \Lambda \, e^{-\kappa
    \sigma/2} \right] \, , \label{Etot}
\ee
where the overall factor of $e^{-\kappa\sigma}$ comes from the Weyl
rescaling to the Einstein frame of the 4D volume factor.  A non-trivial
gradient energy in the dilaton always gives a positive contribution to
the total energy, whereas the 4D potential energy in terms of the
dilaton and volume breathing modes is:
\be
{\mathcal E} = -\frac{4 \pi a_-^2}{\kappa^2} \, e^{-\kappa\sigma}
\frac{1}{2 a_-^2} \left(1 - \frac{a_-^2}{a_i^2} \right)
\ee
Plugging the endpoint configuration
(\ref{endsolution}) into (\ref{Etot}) shows that the potential energy is
negative, $-\frac{4 \pi a_-^2}{\kappa^2}
\, \frac{8 }{ L^2\,z^4}$,   and beats the gradient energy, $\frac{4 \pi
  a_-^2}{\kappa^2}
\, \frac{4 }{ L^2\,z^4}$, giving in total:
\be
{\mathcal E}_{TOT} = - \frac{4 \pi a_-^2}{\kappa^2} \frac{4 \, }{L^2\,z^4}
\ee
This result should be compared to the initial total energy, for which
both the 4D gradient and potential contributions are zero, and so the
energy has been lowered.

\subsubsection{The Decay of Warped Configurations}

As a second example we extend the above analysis from
sphere-monopole compactifications to the more generic presence of
warping, as is required if the two brane tensions are unequal. We
know that configurations with monopole numbers $|N_\I| \geq 2$ are
also unstable in warped brane-world compactifications with
positive-tension brane sources \cite{prs2}. We now seek the
endpoint of this stability, as indicated by the above
oxidation/reduction technique.

To do so we first display a warped solution to the
$(n+6)$-dimensional EYM system with cosmological constant, which
reduces to the warped Minkowski solution of the 6D EYM$\sigma$
theory. As previously, the instability of the 6D solution is
shared by its higher dimensional representation.\footnote{Note,
however, that the direct linearized analysis made in \cite{prs2}
does not extend to warped solutions in (6+n)D because the bilinear
action for the modes of relevance depends on (n+4) in the warped
case \cite{seif}.} We identify a plausible endpoint in the
higher-dimensional EYM theory, and reduce it to identify the
corresponding candidate endpoint in 6D supergravity.

\subsubsection*{The higher-dimensional warped solution}

We again start from the $(n+6)$ dimensional EYM action
(\ref{Daction})
\be
 S= \frac{1}{\kappa_\D^2} \int \exd^\D x \sqrt{-g_\D}
 \left[ R_{(\D)} - \frac{\kappa_\D^2}{4} \, \Tr
 \, {\mathcal F}^2  - \Lambda \right] \,.
\ee
The relevant static warped solution to the corresponding field
equations is obtained by a Weyl rotation of a known black-hole
like solution \cite{bnqtz}, as was done in \cite{ocho} (a similar
solution and method were also used in \cite{msyk}). The result is
\bea \label{SolD}
 \exd s^2_\D &=& r^2 \, g_{\A\B} \exd x^\A \exd x^\B
 + \frac{\exd r^2}{h_\D(r)} + \epsilon_\D^2 \, h_\D(r)\, d\phi^2  \\
 \hbox{and} \qquad  {\mathcal F}_{r\phi} &=& -
 \frac{\epsilon_\D q_{\D}^a Q^a}{r^{n+4}} \, ,\label{SolF}
\eea
where $g_{\A\B}$ is a $d = (n+4)$-dimensional, maximally symmetric
metric, with $R_{\A\B} = (d-1) \lambda_d \, g_{\A\B}$, and we have
introduced an additional parameter, $\epsilon_\D$, which will allow us to
reach the unwarped compactifications {\it via} a smooth limit.  The
function $h_\D(r)$ is given explicitly by
\be\label{Solh}
 h_\D(r) = \lambda_d + \frac{M}{r^{n+3}}
 - \frac{\Lambda r^2}{(n+4)(n+5)}
 - \frac{\kappa_\D^2 q_{\D}^2}{2(n+4)(n+3) \, r^{2(n+3)}} \,,
\ee
where $M$ is an integration constant that can be positive or
negative. This solution can also be found by solving directly the
equations of motion, for which the Einstein equations reduce under
the above ansatz to
\bea\label{Deqs}
 && R_{\A\B}  = - g_{\A\B} \left[ -\frac{(n+3)\lambda_d}{r^2}
 + \frac{h_\D'}{r} + \frac{(n+3) \,h_\D}{r^2}\right]
    = - g_{\A\B} \left[ \frac{\kappa_\D^2\, q_\D^2}{8\,r^{2(n+4)}}
    - \frac{\Lambda}{4}\right] \\
                            \nonumber   \\
 && R_{mn}  = -\frac{1}{2} g_{mn} \left[ h_\D''
 +\frac{(n+4)\,h_\D'}{r} \right] =  g_{mn} \left[
 \frac{3\,\kappa_\D^2 q_\D^2}{8\,r^{2(n+4)}}
 + \frac{\Lambda}{4}\right] \,.
\eea
Ref.~\cite{msyk} shows that this yields the desired warped 6D
solutions to Nishino-Sezgin supergravity found in \cite{ocho} once
the limit $n \to \infty$ is taken, making them a good starting
point for seeking the endpoint of the decay of the underlying
monopole.

The solution above (\ref{SolD}, \ref{Solh}) is very similar to
that studied in \cite{ocho}, Section 3 (see also \cite{GGP,bqtz}).
The geometry is well defined in the region where the metric
function $h_\D(r)$ is positive, and this implies $M > 0$ when
$\lambda_d \le 0$, while $M$ can be negative for $\lambda_d > 0$.
Similar to what is shown in \cite{ocho}, the geometry pinches off
at the points where $h_\D(r)$ vanishes. There are two such real
roots, $r_{\pm}$, since $h_\D(r) \to -\infty$ as $r \rightarrow 0$
and $r \rightarrow \infty$, and changes sign only twice.

Moreover, because $h_\D$ vanishes linearly near $r = r_{\pm}$, being
well approximated by $h_\D(r) \sim h_\D' \left(r_{\pm}\right)
\left(r-r_{\pm} \right)$, the 2D internal metric is approximately
conical at these points, with:
\be
 \exd s_2^2 \sim \exd R_{\pm}^2
 + \left(1- \frac{\delta_{\pm}}{2\pi} \right)^2
 R_{\pm}^2 \exd\phi^2 \,.
\ee
Here $R_{\pm} = 2 \sqrt{(r-r_{\pm})/h_\D'(r_{\pm})}$, and the deficit
angles are given by:
\be \label{deficit}
 \frac{\delta_{\pm}}{2\pi} = 1 - \frac12 \, \vline \,
 \epsilon_\D \, h_\D'(r_{\pm})\, \vline \,.
\ee
These singularities are sourced by codimension-two branes, with
actions
\be
 S_{brane} = -{\mathcal T}_{\pm} \int d^{D-2}y
 \sqrt{-\gamma_{\pm}} \,,
\ee
and whose tensions satisfy $\kappa_\D^2 {\mathcal T}_{\pm} = 2
\delta_{\pm}$. On reduction to the 6D theory these become 3-branes
with tensions $T_{\pm}$ given by ${\mathcal T}_{\pm} V$.

Finally, since the internal space is compact, there is as usual a Dirac
quantization condition for the magnetic flux. Covering the space
with two patches that respectively incorporate $r_{\pm}$,  {\it
\`a la} Wu and Yang, allows the gauge potential to be written
\be
 \mathcal{A}_{\pm} = \frac{\epsilon_\D q_\D^a Q_a}{(n+3)}
 \left(\frac{1}{r^{n+3}} - \frac{1}{r_{\pm}^{n+3}}\right)
 \exd\phi \,.
\ee
These are related by a single-valued gauge transformation on the
overlap only if
\be
 - g \, e_{a\I} \frac{\epsilon_\D q_\D^a}{n+3}
 \left( \frac{1}{r_+^{n+3}}
 -\frac{1}{r_-^{n+3}}\right) = N_\I \,,
 \label{DQ}
\ee
where $e_{a\I}$ are the adjoint charges discussed in earlier
sections, and $N_\I$ is an integer.

In order to have an expression for $h$ in terms of the two real
roots, we can write it as follows ($\ell:= n+3)$:
\bea\label{h1}
 h_\D(r) &=& \lambda_d \left[1-\frac{r_+^\ell}{r^\ell}
 \right]
                \left[1-\frac{r_-^\ell}{r^\ell} \right] \\
&& + \frac{\Lambda}{(\ell+1)(\ell+2)} \frac{1}{r_+^2(r_+^\ell
    -r_-^\ell)}\frac{1}{r^{2\ell}} \left[r^\ell (r_+^{2\,\ell+2}
    -r_-^{2\, \ell+2} )
    -r^{2\, \ell+2} (r_+^\ell -r_-^\ell) \right. \nonumber \\
        && \phantom{00000000000000000000000000000000000000000} \left. -
          (r_+r_-)^\ell  (r_+^{\ell+2} -r_-^{\ell+2} )\right] \nonumber
\eea
where now it is clear that $h(r_\pm) =0$.
By comparing (\ref{Solh} and \ref{h1}), the parameters $r_{\pm}$ can
be related to the original
parameters of the solution as:
\bea
 M &=& \frac{\Lambda}{(\ell+1)(\ell+2)}\frac{(r_+^{2(\ell+1)}
 -r_-^{2(\ell+1)}
  )}{r_+^2 (r_+^{\ell}  -r_-^{\ell})}
                -\lambda_d (r_+^{\ell}  + r_-^{\ell} )
                                \label{M}  \\
 \frac{\kappa_\D^2 q_\D^2}{2\ell(\ell+1)} &=& (r_+r_-)^\ell
            \left[ \frac{\Lambda}{(\ell+1)(\ell+2)}\frac{(
                            r_+^{\ell+2}
                            -r_-^{\ell+2})}{r_+^2 (r_+^{\ell}  -r_-^{\ell})}
                          -\lambda_d\right] \label{q^a}
\eea
and moreover $r_+ = 1$, which amounts to a choice of coordinates.
Meanwhile, the tensions of the branes can be related
to the bulk parameters {\it via} Eq. (\ref{deficit}):
\bea
&& 1- \frac{{\mathcal T}_{+} \kappa_\D^2}{4\pi} =  \label{T_+} \\ &&\frac{1}{2}
    \, \vline  \, \frac{\epsilon_\D}{ r_-^\ell-1}\left((\ell+2 -2\,
      (\ell+1)\, r_-^\ell
      + \ell \, r_-^{2\, \ell+2})\frac{\Lambda}{(\ell+1)(\ell+2)} - \ell (
      r_-^\ell-1)^2 \lambda \right) \vline  \nonumber\\
&& 1- \frac{{\mathcal T}_{-} \kappa_\D^2}{4\pi}= \label{T_-} \\ && \frac{1}{2}
    \, \vline  \, \frac{\epsilon_\D}{r_-^{\ell+1}(r_-^\ell-1)} \left((-\ell
      +2\, (\ell+1)\, r_-^{\ell+2}
      -(\ell+2) \, r_-^{2\, \ell+2})\frac{\Lambda}{(\ell+1)(\ell+2)}
     + \ell (
      r_-^\ell-1)^2 \lambda \right)\vline \, .\nonumber
\eea

\bigskip

\noindent
{\it Unwarped limit}

\bigskip

\noindent
As an aside, we show how the above higher dimensional warped background
reduces to the known
unwarped solution as the warp factor goes to one, $r_-
\rightarrow r_+$.  This limit can be taken by making the change of
coordinates:
\be
 r=\frac{r_+}{2} \left((1+\alpha) +
 (1-\alpha)\cos\theta \right)\, ,
\ee
where we have defined $\alpha := r_-/r_+$.  We then take $\alpha =
1+\xi$ together with the limit $\xi \rightarrow 0$, but insist that
$\xi \, \epsilon_\D \rightarrow \varepsilon_\D$, a finite constant.  In this
way, the metric assumes the form of the rugbyball
\be
 ds_D^2 = g_{AB} dx^A dx^B + a^2 \left( d\theta^2 + \beta^2
  \sin^2\theta d\phi^2 \right) \, ,
\ee
where the radius and deficit angle are, respectively,
\bea
 a^2 &=& \frac{1}{\Lambda-\ell^2\lambda} \qquad \rm{and} \label{eom1}\\
 \beta^2 &=& \frac{\varepsilon_\D^2}{4 a^4} \,,
\eea
and the gauge field is that of the monopole
\be
 {\mathcal A}_{\pm} = \frac{\varepsilon_\D q_\D^a Q_a}{2}
 \left(\cos\theta \mp 1
 \right) \, \exd\phi \, .
\ee
The quantisation
condition (\ref{DQ}) reduces to $- g \, \varepsilon_\D \, q_\D^a\,
e_{a\I}=N_\I$, and   Eq. (\ref{q^a}) tells us that:
\be
 \ell \lambda = \frac{1}{(\ell+1)}\left[\Lambda -
 \frac{\kappa_\D^2 \varepsilon_\D^2 q_\D^2}{8\beta^2 a^4} \right] \, ,\label{eom2}
\ee
which is precisely one of the constraints encountered from the
equations of motion for the unwarped rugbyball.
Meanwhile, the boundary conditions (\ref{T_+}, \ref{T_-}) also reduce to
the expected ones:
\bea
 \frac{{\mathcal T}_{+} \kappa_\D^2}{4\pi}&=& 1-\beta \\
 \frac{{\mathcal T}_{-} \kappa_\D^2}{4\pi} &=& 1-\beta \, .
\eea
We obtain an additional constraint by putting together Eqs.
(\ref{q^a}) and (\ref{T_+}, \ref{T_-}): after some manipulation
one arrives at the condition $\lambda=0$.  Therefore, we are able
to take the unwarped limit only for flat $(n+4)$D slices, and the warped generalizations for the
$\hbox{dS}_{4+n} \times S_2$ and $\hbox{AdS}_{4+n} \times S_2$
solutions are yet to be discovered. Finally, taking furthermore $\beta \rightarrow 1$ we
recover the equations for the sphere (\ref{DSol1},\ref{radios2}).

\subsubsection*{The Warped Endpoints: Upstairs and Downstairs}

Back to the main line of argument, having established the higher
dimensional warped configurations that are assumed in the presence
of 3-branes, we now ask what happens to these configurations when
they are unstable.  As described above, we expect both the 6D and
$D$-dimensional warped solutions to be unstable when there are
monopoles numbers $|N_\I| >1$. The monopole numbers depend on the
charges present, as well as the parameters $q_\D^a, \epsilon_\D$
and $r_-$, {\it via} the Dirac Quantisation condition (\ref{DQ}).
The $d=4+n$-dimensional curvature, $\lambda_d$, completes the
description of the solution (\ref{SolD}-\ref{Solh}), but not all
of these parameters are independent, due to the equations of
motion (\ref{q^a}, \ref{T_+}, \ref{T_-}).  Thus we can specify a
given solution completely with one parameter, say, $q_\D^a$.
Beginning with an unstable solution, $q_{\D \,i}^a$, the monopole
field strength will decay conserving its topological flux, as we
have seen previously, and the geometry will adjust appropriately
according to the equations of motion.  A reasonable endpoint in
the $D$-dimensional EYM theory is then a warped configuration
within the same class (\ref{SolD}-\ref{Solh}), with new parameters
$q_{\D \,f}^a, \epsilon_{\D \, f}, r_{-_f}$ and $\lambda_{d_f}$.

It remains to play the same game as in unwarped case to discover
how the geometry and dilaton respond to the decay of the monopole
in 6D supergravity.  The rules of the game are by now familiar; we
begin with a warped dilatonic 6D model, tuning the bulk
cosmological constant, $\Lambda$, to allow for flat 4D slices in
the initial unstable configuration, with monopole strength
$q^a=q^a_i$ and brane tensions $T_{\pm}$. Uplifting this model to
a non-dilatonic D-dimensional theory, it is easy to see that the
decay of the monopole to its stable topological cousin curves the
$(4+n)$D slices from $\lambda_{d_i} = 0$ to $\lambda_{d_f}
\neq 0$. Now we dimensionally reduce the stable D-dimensional
solution, and take the $n \rightarrow \infty$ limit, in order to
recover the geometry and dilaton profile in the 6D supergravity
model.

The dimensional reduction is performed as in the previous section.
To allow a well-defined $n\rightarrow \infty$ limit, we further
make the change of coordinates $\rho = r^{2+\frac{n}{2}}$, along
with the parameter redefinitions $\rho_- = r_-^{2+\frac{n}{2}}$,
$\epsilon = 2\, \epsilon_\D/n$ and $\lambda = n^2
\lambda_d$.  Moreover, we define the function $h(\rho)$ as
$h(\rho) = \lim_{n\rightarrow \infty} \frac{n^2 \rho}{4}\,
h_n(\rho)$:
\be
 h(\rho) = \left(\lambda -\Lambda
 \right)\frac{\rho}{4}
 \left[1-\frac{1}{\rho^2}\right]\left[1-\frac{\rho_-^2}{\rho^2}\right]
  \, . \label{tildeh}
\ee
Finally, the solution to the 6D supergravity can be written:
\bea
 &&ds_6^2 = \rho \, dz^2 + \rho \, \left(\frac{z}{L}\right)^2 \left(-dt^2 +
  \delta_{ij} dx^i dx^j
 \right) + \left(\frac{z}{L}\right)^2 \left( \frac{d\rho^2}{h(\rho)} +
  \epsilon^2 h(\rho) d\phi^2 \right) \nonumber \\
 &&\kappa\sigma = 4 \ln \left(\frac{z}{L}\right) + 2 \ln \rho \qquad \qquad F_{\rho\phi} = -
 \frac{\epsilon \, q^a}{\rho^3}Q_a \label{warpedendpoint}
\eea
where, assuming $\lambda <0$, we have defined $z$  as
in Eq. (\ref{z}), $4/L =\sqrt{-\lambda} $ and the
quantization condition takes the familiar form:
\be
 - g \, e_{a\I} \frac{\epsilon \, q^a}{2}
 \left( 1
 -\frac{1}{\rho_-^{2}}\right) = N_\I \,,
\ee
where we recall the choice of coordinates such that $\rho_+ = 1$.
The parameters describing the background, $\lambda, \epsilon$ and
$\rho_-$, are given as above in terms of $q_f,
T_{\pm}$, using (\ref{q^a},\ref{T_+},\ref{T_-}) in the limit
$n\rightarrow \infty$:
\bea
 \lambda &=& \Lambda-\frac{\kappa^2
 q_f^2}{2\rho_-^2}
 \nonumber \\
 \left(1-\frac{T_{+} \kappa^2}{4\pi}\right)^2 &=& \left(\frac{\epsilon}{4}
    \frac{1}{(\rho_-^2-1)}\left((1 -2\rho_-^2
      + \rho_-^4)\Lambda - (\rho_-^2-1)^2 \lambda \right)\right)^2
    \nonumber \\
 \left(1-\frac{T_{-} \kappa^2}{4\pi}\right)^2 &=& \left(\frac{\epsilon}{4}
    \frac{1}{\rho_-^2(\rho_-^2-1)}\left((-1 +2\rho_-^2
      - \rho_-^4)\Lambda + (\rho_-^2-1)^2 \lambda \right)\right)^2 \, .
    \label{parameters}
\eea
It is a simple exercise to invert these expressions.  Then, the
initial tuning of $\Lambda$ gives:
\be
 \Lambda=\frac{\kappa^2 q_i^2}{2} \frac{\left(1-\frac{T_+
 \kappa^2}{4\pi}\right)}{\left(1-\frac{T_-
    \kappa^2}{4\pi}\right)}
\ee
whereas the final solution parameters are:
\bea
 \lambda &=& \Lambda - \frac{\kappa^2 q_f^2}{2} \frac{\left(1-\frac{T_+
 \kappa^2}{4\pi}\right)}{\left(1-\frac{T_-
    \kappa^2}{4\pi}\right)}
 \nonumber \\
 \epsilon &=& \pm \frac{32\pi}{\kappa^2 q_f^2}  \frac{\left(1-\frac{T_+
 \kappa^2}{4\pi}\right)^2}{\left(T_+ \kappa^2- T_-
    \kappa^2\right)} \nonumber \\
 \rho_- &=& \sqrt{\frac{1-\frac{T_+ \kappa^2}{4\pi}}{1-\frac{T_-
      \kappa^2}{4\pi}}} \, .
\eea
In contrast to the unwarped case with or without branes, here we
find a unique physical solution with $\lambda <0$ (for the monopole
field strength to decay, the combination $\epsilon \, q^a$ must decrease,
which, together with $\epsilon \sim 1/q^2$, implies $q_f^2 > q_i^2$).
Otherwise, the endpoint in the presence of warping is a
straightforward generalization to the one we found in the previous
sections, breaking the 4D maximal symmetry, and it is similarly the analogue of the warped scaling
solutions found in \cite{scaling}.
Although to establish the
stability of this final solution would now require a systematic
analysis of its fluctuations, we argue that due to flux
conservation, the monopole has nowhere else to go.

\subsubsection*{Energy}

Moreover, we now confirm that the energy of our proposed endpoint solution is
less than the zero energy of the initial unstable
configuration.  The total energy density can be defined as in the
unwarped case
as a sum of 4D gradient and potential energies, (\ref{Etot}),
but now including the warp factor when we integrate out the extra
dimensions.  Evaluating on the background solution
(\ref{warpedendpoint}), we find, just as for the unwarped case, that
the gradient energy is $-1/2$ of the potential energy, so that the
total energy in the Einstein frame ($g^{(E)}_{\mu\nu} = (z/L)^2
g_{\mu\nu}$) is:
\be
{\mathcal E}_{TOT} = - \frac{4 \pi \epsilon}{\kappa^2}
\frac{(1-\rho_-^2)}{L^2\,z^4}
\ee
and indeed less than zero.

\section{Conclusions}

Compactifications supported by gauge field fluxes were long ago
\cite{seifinstabilities, schellekens} found
to be
generically unstable, due to tachyonic modes in the non-Abelian degrees of
freedom, but the fate that they meet has remained an open question.
In this paper, we have explored a number of possible candidates for the endpoint of
this instability.

Topological flux conservation suggests that an unstable monopole field
decays to the unique, topologically connected, stable monopole
\cite{brandtneri1, brandtneri2, coleman}, and we have
determined how the geometry responds to this decay in various scenarios.  In
the Einstein-Yang Mills theory, with a cosmological constant, a
$\hbox{Mink}_d \times S_2$ lowers its potential energy by adjusting to
$\hbox{AdS}_d \times S'_2$.  In 6D supergravity, the dilaton precludes such a
simple dynamics, and we have argued that it forces the breaking of the maximal
symmetry in the non-compact dimensions.  By recasting the dilaton as the
volume modulus of $n$ fictitious dimensions in a yet-higher dimensional
non-dilatonic Einstein-Yang Mills theory \cite{tanaka}, we were able
to find the corresponding solutions explicitly for both unwarped and warped
initial configurations, with and without brane sources.  The non-trivial profile of the
dilaton in 4D generates a singular, static, Kasner-like geometry that
is conformal to (unwarped or warped)
$\hbox{Mink}_4 \times S_2$, 
where the radius of the 2-sphere grows with the distance from the singularity.  
How to interpret the naked timelike singularity
to which the
instability seems to lead is an important open question; does it
signal an inconsistency or does it suggest some 
new physics beyond any supergravity approximation?  One way to resolve
the singularity is to pass to the 
higher dimensional Einstein-Yang Mills theory in (6+$n$)D, in which case the singularity results from projecting the smooth $\hbox{AdS}_{4+n} \times S_2$ geometry onto six dimensions.  Such ideas have been discussed in \cite{dilatonicbh}.  Moreover, we have shown that the final
configuration is perturbatively stable, and that the decay results in
a finite total energy which is lower (counting gradient and potential
contributions) than the initial one.

We would like to end with a comment.  The instability suffered by
Yang-Mills sectors in the background of a monopole is the
spherical analogue of the Nielsen-Olesen instability that occurs
in flat 4D Yang-Mills theory \cite{qcd}.  In that case, it was
proposed that condensation of the tachyonic modes leads to the
formation of magnetic flux tubes \cite{ambjorn}, in a rather
beautiful imitation of the vortex formation in superconductor
physics \cite{weinberg}. That such a dynamics might also be
possible in the present case is certainly an interesting
speculation.

\section*{Acknowledgements}
We would like to thank Hyun-Min Lee and Gianmassimo Tasinato for
many helpful conversations and insights about extra-dimensional
dynamics. CB is supported in part by funds
from the Natural Sciences and Engineering Research Council (NSERC)
of Canada, CERN and McMaster University. Research at the Perimeter
Institute is supported in part by the Government of Canada through
NSERC and by the Province of Ontario through MRI. SLP is supported by the Deutsche Forschungsgemeinschaft
(DFG) under the Collaborative Research Center (SFB) 676, by the
European Union 6th framework program MRTN-CT-503359 "Quest for
Unification". IZ is partially supported by the European Union 6th
framework program MRTN-CT-2006-035863 "UniverseNet" and
SFB-Transregio 33 "The Dark Universe" by the DFG.

\end{document}